\documentclass[12pt,draftclsnofoot,peerreviewca,letterpaper,onecolumn]{IEEEtran}

\usepackage{cite}
\usepackage{graphicx,color,epsfig,rotating}
\usepackage{amsfonts,amsmath,amssymb,bbm}
\usepackage{algorithm}
\usepackage{algpseudocode}
\usepackage{amsmath}
\usepackage{cite}
\usepackage{mdwtab}
\usepackage{placeins}
\usepackage{psfrag, graphicx}
\usepackage[latin1]{inputenc}
\usepackage{amssymb}
\usepackage{makeidx}
\usepackage{epstopdf}
\usepackage{enumitem}
\usepackage{amssymb}
\usepackage{cite}
\usepackage{graphicx,color,epsfig,rotating}
\usepackage{amsfonts,amsmath,amssymb,bbm}
\usepackage{algorithm}% http://ctan.org/pkg/algorithms
\usepackage{algpseudocode}% http://ctan.org/pkg/algorithmicx
\usepackage{subcaption}
\usepackage{amsmath}
\usepackage{cite}
\usepackage{placeins}
\usepackage{graphicx}
\usepackage[latin1]{inputenc}
\usepackage{amssymb}
\usepackage{multirow}
\usepackage{stfloats}
\usepackage{tabularx}
\usepackage{booktabs}
\usepackage{url}
\usepackage{bm}
\usepackage{soul}
\usepackage{pstricks}

\newfont{\bbb}{msbm10 scaled 700}

\newfont{\bb}{msbm10 scaled 1100}

\newcommand{\gv}{{\bf g}}

\newcommand{\uv}{{\bf u}}

\newcommand{\vv}{{\bf v}}

\newcommand{\yv}{{\bf y}}

\newcommand{\eqdef}{\stackrel{\Delta}{=}}

\newcommand{\be}{\begin{equation}}
\newcommand{\ee}{\end{equation}}
\newcommand{\bea}{\begin{eqnarray}}
\newcommand{\eea}{\end{eqnarray}}

\usepackage{changes}
\definechangesauthor[color=blue,name={Mingyue Ji}]{JMY}
\UseRawInputEncoding
\newtheorem{example}{Example}%[section]
\newtheorem{theorem}{Theorem}%[section]
\newtheorem{corollary}{Corollary}%[section]
\newtheorem{remark}{Remark}%[section]
\newtheorem{observation}{Observation}

\ifodd 1

\else

\fi

\ifodd 1

\else

\fi

\begin{document}
\setcounter{page}{1}
\title{%Cache-aided Interference Alignment for
On the Fundamental Limits of Cache-aided Multiuser Private Information Retrieval}
\author{Xiang~Zhang,~\IEEEmembership{Student Member,~IEEE},
        Kai~Wan,~\IEEEmembership{Member,~IEEE}, \\
        Hua~Sun,~\IEEEmembership{Member,~IEEE},
        Mingyue Ji,~\IEEEmembership{Member,~IEEE}, \\
        and~Giuseppe~Caire,~\IEEEmembership{Fellow,~IEEE}
%\thanks{Part of this work was presented in the CCDWN workshop of WiOpt 2020 conference, Volos, Greece.}
\thanks{This manuscript was partially presented in the conference papers \cite{zhang2020isit,zhang2020private}.}
\thanks{X. Zhang and M. Ji are with the Department of Electrical and Computer Engineering, University of Utah, Salt Lake City, UT, 84112, USA (E-mail:~\{xiang.zhang, mingyue.ji\}@utah.edu).}
\thanks{K. Wan and G. Caire are with the Electrical Engineering and Computer Science Department, Technical University of Berlin,
10587, Berlin, Germany (E-mail: \{kai.wan, caire\}@tu-berlin.de).}
\thanks{H. Sun is with the Department of Electrical Engineering, University of North Texas, Denton, TX, 76203 (E-mail:
hua.sun@unt.edu).}
}

\maketitle
\begin{abstract}
We consider the problem of cache-aided Multiuser Private Information Retrieval (MuPIR) which is an extension of the single-user cache-aided PIR problem to the case of multiple users. In MuPIR,
each of the $K_{\rm u}$ cache-equipped users wishes to privately retrieve
a message out of $K$
messages from $N$
databases each having access to the entire message library. The privacy constraint requires that any individual database learns nothing about the demands of all users. The users are connected to each database via an error-free shared-link.
In this paper,
we aim to characterize the optimal trade-off between users' memory and communication load for such systems.
Based on the proposed novel approach of \emph{cache-aided interference alignment (CIA)}, first, for
the MuPIR problem with $K=2$ messages, $K_{\rm u}=2$ users and $N\ge 2$ databases, we propose achievable retrieval schemes for both uncoded and general cache placement.
The CIA approach is optimal when the cache placement is uncoded. For general cache placement, the CIA approach is optimal  when $N=2$ and $3$ verified by the computer-aided approach.
Second,
when $K,K_{\rm u}$ and $N$ are general, we propose a new \emph{product design} (PD) which incorporates the PIR code into the linear caching code.
 The product design is shown to be order optimal within a multiplicative factor of 8 and is exactly optimal when the user cache memory size is large.
\end{abstract}
%--------------------------------------------------------------
\section{Introduction}
\label{section: introduction}
Introduced by Chor \emph{et al.} in 1995 \cite{chor1995private}, the problem of private information retrieval (PIR) seeks efficient ways for a user to retrieve a desired message from  $N$  databases, each holding a library of $K$ messages, while keeping the desired message's identity private from each database. Sun and Jafar (SJ) recently characterized the capacity of the PIR problem with non-colluding databases  \cite{blindPIR,sun2017capacity}.
{Coded caching was originally proposed by Maddah-Ali and Niesen (MAN) in~\cite{maddah2014fundamental} for a shared-link caching network consisting of a server, which is connected to  $K_{\rm u}$ users through a noiseless broadcast channel and has access to a library of $K$ equal-length files.
Each user can cache
$M$ files and demands one file. The MAN scheme
proposed a combinatorial
cache placement design so that during the delivery phase, each
transmitted coded
message is simultaneously useful to
multiple users such that the communication load can be significantly reduced. Under the constraint of uncoded cache placement and for worst-case load, the MAN scheme was proved to be optimal when $K \geq K_{\rm u}$~\cite{wan2016optimalityTIT} and order optimal within a factor of $2$ in general \cite{yu2018characterizing}.

The combination of privacy and coded caching has gained significant
attentions recently. Two different privacy models are commonly
considered. First,
in \cite{tandon2017capacity,wei2018fundamental,8362308}, the \emph{user-against-database} privacy model was studied where individual databases are prevented from learning the {single-user's demand}. The author in \cite{tandon2017capacity} studied the case where a single cache-aided user is connected to a set of $N$ replicated databases and showed that memory sharing between the memory-load pairs $(0,1+\frac{1}{N}+\cdots+\frac{1}{N^{K-1}})$ and $(K,0)$ (i.e., split the messages and cache memory proportionally and then implement
two PIR schemes on the independent parts of the messages) is actually optimal if the databases are aware of the user's cached content. However, if the databases are unaware of the user's cached content, then there is a multiplicative ``unawareness gain" in capacity in terms of the user memory as shown in \cite{wei2018fundamental,8362308}.
Second, the authors in \cite{wan2019coded,wan2019device,kamath2019demand,sarvepalli2019subpacketization} considered the \emph{user-against-user} privacy model where users are prevented from learning each other's demands. The authors in \cite{wan2019coded} first formulated the \emph{coded caching with private demands} problem where a shared-link cache network with demand privacy, i.e., any user can not learn anything about the demands of other users, was considered. The goal is to design efficient delivery schemes such that the communication load is minimized while preserving such privacy. Order optimal schemes were proposed based on the  concept of virtual user.

This paper formulates the problem of cache-aided Multiuser PIR (MuPIR), where each of the $K_{\rm u}$ cache-equipped users aims to retrieve a message from  $N$ distributed non-colluding databases while preventing any one of them from gaining knowledge about the user demands given that the cached content of the users are known to the databases.\footnote{\label{foot:cannot user private caching} {Note that the virtual user strategy and the strategy based on scalar linear function retrieval for coded caching with private demands \cite{wan2019coded,arxivfunctionretrieval,yan2020private}  were designed
based on the fact that the user caches are not transparent to each other. Therefore, they cannot be used
in the considered MuPIR problem, where databases are aware of the user caches.}}
The  contributions of this paper are as follows.
First, based on the novel idea of {\em cache-aided  interference alignment (CIA)}, we construct cache placement and private delivery phases achieving the
memory-load pairs $\left( \frac{N-1}{2N},\frac{N+1}{N}\right)$ and $\left( \frac{2(N-1)}{2N-1},\frac{N+1}{2N-1}\right)$ for the MuPIR problem with $K=2$ messages, $K_{\rm u}=2$ users and $N\ge 2$ databases. Different from the existing cache-aided interference alignment schemes in~\cite{cacheaidedinter,interferencemanagement,degreesHachem2018,zhang2020cache}
which were designed for the cache-aided interference channels, the purpose of our proposed private delivery scheme is to let each server send symmetric messages (in order to keep user demand privacy), each of which contains some uncached and undesired symbols (i.e., interference) of each user.
The proposed CIA approach effectively aligns these interference for each user and thus facilitates correct decoding. We prove that the proposed scheme is optimal under the constraint of uncoded cache placement.
Computer-aided investigation   given in \cite{tph:19:cai} also shows that the proposed schemes are optimal for general cache placement when $N=2 $ and $3$.
Second, for general system parameters $K,K_{\rm u}$ and $N$, we propose a \emph{Product Design (PD)} which incorporates the
SJ scheme \cite{sun2017capacity} into the
MAN coded caching scheme \cite{maddah2014fundamental}. Interestingly, the load of the proposed design
is the product of the loads achieved by these two schemes and is optimal within a factor of 8.
Moreover, the PD is exactly optimal when the users' cache memory size is beyond a certain threshold.
Finally,
we characterize the optimal memory-load trade-off for the case of $K=K_{\rm u}=N=2$ where users request distinct messages. It is shown that under the constraint of the distinct demands, the achieved load can be smaller than the case without   constraint of distinct demands.

The paper is organized as follows. In Section~\ref{sec: problem-formulation}, we present the formal problem formulation. The main results of this paper are given in Section~\ref{sec: Main Result}. In Section~\ref{Section: achievability}, we describe the proposed CIA schemes in detail and in Section~\ref{sec: theorem 4}, we present the product design for general system parameters. We discuss some interesting observations for distinct demands in Section~\ref{sec: Discussions}. Finally, we conclude this paper and present several future directions
in Section~\ref{sec: conclusions}.

\paragraph*{Notation Convention}
$|\cdot|$ represents the cardinality of a set. $[n]\triangleq\{1,2,\cdots,n-1,n\}$, $[m:n]\triangleq\{m,m+1,m+2,\cdots,n\}$ and $(m:n)\triangleq(m,m+1,\cdots,n)$ for some integers $m\le n$. $\mathbb{Z}^+$ denotes the set of non-negative integers. For two sets $\mathcal{A}$ and $\mathcal{B}$, define the difference set as $\mathcal{A}\backslash \mathcal{B}\triangleq\{x\in\mathcal{A}:x\notin \mathcal{B}\}$. For an index set $\mathcal{I}$, the notation $A_{\mathcal{I}}$ represents the set $\{A_i:i\in\mathcal{I}\}$. When $\mathcal{I}=[m:n]$, we write $A_{[m:n]}$ as $A_{m:n}$ for brevity. For an index vector $I=(i_1,i_2,\cdots,i_n)$, the notation $A_I$ represents a new vector $A_I\triangleq(A_{i_1},A_{i_2},\cdots,A_{i_n})$. $\bm{0}_n$ denotes the all-zero vector with length $n$, i.e., $\bm{0}_n\triangleq(0,0,0,\cdots,0)$.
Let $\mathbf{1}_n\triangleq(1,1,...,1)$ with length $n$. $\mathbf{I}_n$ denotes the $n\times n$ identity matrix. For a matrix $\mathbf{A}$, $\mathbf{A}(i,:)$ and $\mathbf{A}(:,j)$ denote the $i$-th row and $j$-th column of $\mathbf{A}$ respectively. $\mathbf{A}^{\rm T}$ represents the transpose of $\mathbf{A}$. $\mathbb{F}_2$ represents the binary field.
In this paper, the operations (addition and linear combination) are on the binary field.

%%========================================================================================================
\section{Problem Formulation}
\label{sec: problem-formulation}
\begin{figure}
\centering
\includegraphics[width=0.42\textwidth]{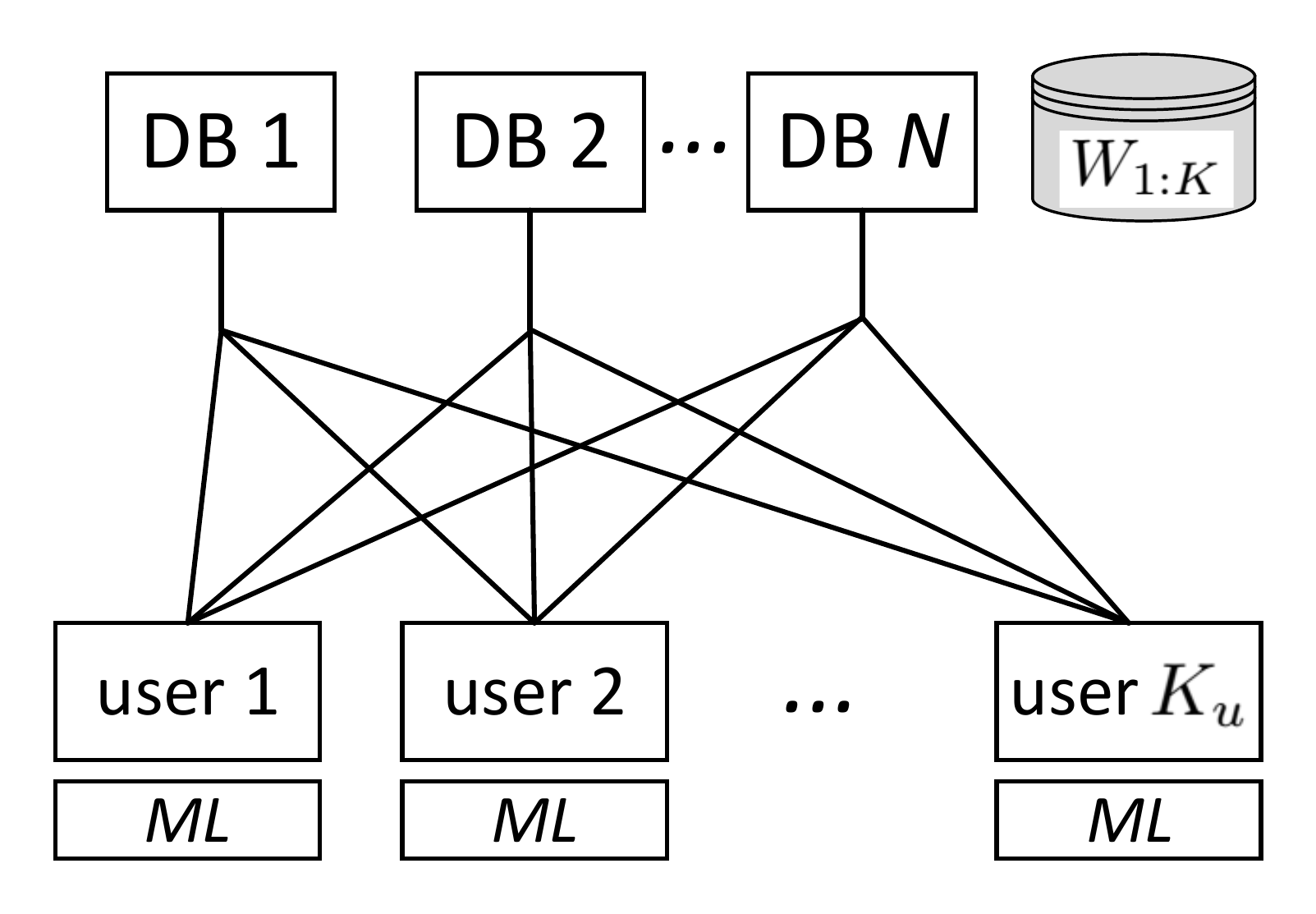}
\vspace{-0.5cm}
\caption{\small Cache-aided MuPIR system with $N$ replicated databases, $K$ independent messages and $K_{\rm u}$ cache-equipped users. The users are connected to each DB via an error-free shared-link broadcast channel.}
\label{fig: sys_model}
\vspace{-0.8cm}
\end{figure}

We consider a system with $K_{\rm u}$ users, each
aiming to privately retrieve a message from $N\ge 2$ replicated non-colluding databases (DBs). Each DB stores $K$ independent messages, denoted by $W_1,W_2,\cdots,W_K$, each of which is uniformly distributed over $[2^L]$.
Each user is equipped with a cache memory of size $ML$ bits, where $0\le M\le K$. Let the random variables $Z_1,Z_2,\cdots,Z_{K_{\rm u}}$ denote the cached content of the users. The system operates in two phases, the \emph{cache placement phase} followed by the \emph{private delivery phase}. In the cache placement phase, the users fill up their cache memory without knowledge of their future demands.
We assume that the cached content of each user is a
function of
$W_{1:K}$ and is known to all DBs. In the private delivery phase, each user $k\in[K_{\rm u}]$ wishes to retrieve a message $W_{\theta_k}$ where $\theta_k \in [K]$. Let {$\bm{\theta}\triangleq(\theta_1,\theta_2,\cdots,\theta_{K_{\rm u}})$} denote the demands of the users. We assume that $\bm{\theta}$ follows an arbitrary distribution with full support over $[K]^{K_{\rm u}}$. {Depending on $\bm{\theta}$ and $Z_1,Z_2,\cdots,Z_{K_{\rm u}}$, the users cooperatively generate $N$ queries $Q^{[\bm{\theta}]}_{1},Q^{[\bm{\theta}]}_{2},\cdots,Q^{[\bm{\theta}]}_{N}$, and then send the query $Q^{[\bm{\theta}]}_{n}$ to DB $n$.} Upon receiving the query,  DB $n$ responds with an answer {$A^{[\bm{\theta}]}_{n}$ broadcasted to all users.} The answer {$A^{[\bm{\theta}]}_{n}$} is a deterministic function of
{$Q^{[\bm{\theta}]}_{n}$},
$W_{1:K}$ and $Z_{1:K}$, which written in terms of conditional entropy, is
\be
H({A^{[\bm{\theta}]}_{n}}|{Q^{[\bm{\theta}]}_{n}},W_{1:K},Z_{1:K})=0,\quad \forall n\in[N].
\ee
After collecting all the answers from the $N$ DBs, the users
can recover their desired messages error-free using
their cached information, i.e.,
\be \label{eqn: decodabilty}
H(W_{\theta_k}|{Q^{[\bm{\theta}]}_{1:N}},{A^{[\bm{\theta}]}_{1:N}},Z_k)=0, \quad \forall k\in[K_{\rm u}].
\ee
To preserve the privacy with respect to the DBs, it is required that \footnote{{ The privacy constraint (\ref{eqn: privacy}) can be equivalently written as $I(\bm{\theta}; Q^{[\bm{\theta]}}_{n}, W_{1:K},Z_{1:K})=0,\forall n\in[N]$ since the answer $A_n^{[\bm{\theta}]}$ is a deterministic function of $Q_n^{[\bm{\theta}]}$ and $W_{1:K}$, which implies $I(\bm{\theta};{ Q^{[\bm{\theta]}}_{n}, A^{[\bm{\theta}]}_{n}},W_{1:K},Z_{1:K})=I(\bm{\theta}; Q^{[\bm{\theta]}}_{n}, W_{1:K},Z_{1:K}) + I(\bm{\theta}; A^{[\bm{\theta]}}_{n}| Q^{[\bm{\theta]}}_{n} W_{1:K},Z_{1:K} )=I(\bm{\theta}; Q^{[\bm{\theta]}}_{n}, W_{1:K},Z_{1:K})$.     }}
\be \label{eqn: privacy}
I(\bm{\theta};{ Q^{[\bm{\theta]}}_{n}, A^{[\bm{\theta}]}_{n}},W_{1:K},Z_{1:K})=0,\quad \forall n\in[N].
\ee
Let $D$ denote the total number of bits broadcasted from the DBs, then  the  \emph{load} (or transmission rate) of the MuPIR problem is defined as
\be
R \eqdef \frac{D}{L}=\frac{\sum_{n=1}^{N}H({ A^{[\bm{\theta}]}_{n})}}{L}.
\ee
From the privacy constraint~\eqref{eqn: privacy}, the expression of load can be also written as
\be \label{eq:load on individual thetas}
R =\frac{\sum_{n=1}^{N}H({ A^{[\bm{\theta}^i]}_{n})}}{L}, \ \forall i \in \left[ K^{K_{\rm u}}\right],
\ee
where $\bm{\theta}^i$ represents  the  $i$-th realization of all the $K^{K_{\rm u}}$ possible realizations of the demand vector. This is because the load $R$ should not depend on the user demands $\bm{\theta}$ otherwise it leaks information about the user demands to the DBs and (\ref{eqn: privacy}) would not be possible.

A memory-load pair $(M,R)$ is said to be achievable if there exists a MuPIR scheme satisfying the decodability constraint (\ref{eqn: decodabilty}) and the privacy constraint (\ref{eqn: privacy}). The goal of the MuPIR problem is to design the cache placement and the corresponding private delivery phases such that the load is minimized. For any $0\le M\le K$, let $R^{\star}(M)$ denote the minimal achievable load. In addition, if  the users directly cache a subset of the library
bits, the placement phase is said to be {\it uncoded}.  We denote the minimum achievable load under the constraint of uncoded cache placement by
 $R^{\star}_{\textrm{uncoded}}(M)$.

 Note that any converse bound on the worst-case load for the coded caching problem without privacy constraint formulated in~\cite{maddah2014fundamental} is also a converse bound on  $R^{\star}(M)$. Suppose $R^{\prime}(M)$ is a converse bound on the worst-case load for the coded caching problem without privacy. Then for any achievable caching scheme there must be exist one demand vector realization $\bm{\theta}^{\prime}\in[K]^{K_{\rm u}}$ such that in order to satisfy this demand,  the  total number of broadcasted bits is no less than $R^{\prime}(M)L$.
With the consideration of the privacy constraint in~\eqref{eqn: privacy} and the definition of the load in~\eqref{eq:load on individual thetas}, it can be seen that
\be \label{eqn: existing converse}
R^{\star}(M) = \frac{1}{L} \sum_{n=1}^{N}H({ A^{[\bm{\theta}^{\prime}]}_{n})} \geq R^{\prime}(M).
\ee
Similarly, any converse bound on the worst-case load under the constraint of uncoded cache placement for the coded caching problem without privacy constraint formulated in~\cite{maddah2014fundamental} is also a converse bound on  $R^{\star}_{\textrm{uncoded}}(M)$. Suppose $R^{\prime\prime}(M)$ is a converse bound on the worst-case load under the constraint of uncoded cache placement for the coded caching problem without privacy; thus we have
\be \label{eqn: existing uncoded converse}
R^{\star}_{\textrm{uncoded}}(M) \geq R^{\prime\prime}(M).
\ee

 %%=====================================================================================================================
\section{Main Results}
\label{sec: Main Result}
First, we consider the MuPIR peoblem with parameters
$K=K_{\rm u}=2$ and $N\ge 2$. In this case, we propose a novel \emph{cache-aided interference alignment} (CIA) based scheme (see Section \ref{Section: achievability}) and the corresponding load is given by Theorem \ref{theorem 1}.
\begin{theorem}
\label{theorem 1}
For the cache-aided MuPIR problem with $K=2$ messages, $K_{\rm u}=2$ users and $N\ge 2$ DBs, the following load is achievable
\begin{equation}\label{eqn: achiavable rate}
  R_{\textrm{CIA}}(M) =
    \begin{cases}
      2(1-M) ,& 0\le M\le \frac{N-1}{2N}, \\
      \frac{(N+1)\left(3-2M \right)}{2N+1} ,  & \frac{N-1}{2N}\le M\le \frac{2(N-1)}{2N-1}, \\
      \frac{(N+1)(2-M)}{2N}, & \frac{2(N-1)}{2N-1}\le M\le 2.
    \end{cases}
\end{equation}
\end{theorem}
\begin{IEEEproof}
The proof of Theorem \ref{theorem 1} is provided in Section \ref{Section: achievability}, where we
present the CIA approach achieving the memory-load pairs $\left(\frac{N-1}{2N},\frac{N+1}{N} \right)$ and $\left( \frac{2(N-1)}{2N-1} ,\frac{N+1}{2N-1}\right)$. Together with the two trivial pairs $(0,2)$ and $(2,0)$, we obtain four corner points. By memory sharing among these corner points, the load in Theorem \ref{theorem 1} can be achieved.
\end{IEEEproof}

\begin{remark}
The computer-aided approach given in \cite{tph:19:cai} shows that the achievability result in Theorem \ref{theorem 1} is
optimal when $N=2,3$.
For $N\ge 4$, the converse
remains open. In addition,   the achieved load by the CIA  based scheme is strictly better than applying twice the single-user cache-aided PIR of \cite{tandon2017capacity} for each user, which yields a load of
$\frac{(N+1)(2-M)}{N}\ge R_{\textrm{CIA}}(M),\forall M\in[0,2]$ (i.e., the achieved memory-load tradeoff is the memory sharing between $(0,\frac{N+1}{N})$ and $(2,0)$).
\end{remark}

\begin{corollary}
\label{corollary: theorem 1 optimality}
The load $R_{\textrm{CIA}}(M)$ in Theorem \ref{theorem 1} is optimal when
$M\ge \frac{2(N-1)}{2N-1} $.
\end{corollary}
\begin{IEEEproof}
When $M \ge \frac{2(N-1)}{2N-1}$, the load of $R_{\textrm{CIA}}(M)=\frac{(N+1)(2-M)}{2N}$ coincides with the converse bound for the single-user cache-aided PIR in\cite{tandon2017capacity}. Since increasing
the number of users $K_{\rm u}$
cannot
decrease the load, the achieved load in Theorem \ref{theorem 1} is optimal.
\end{IEEEproof}

%---------------------------------------------
\begin{corollary}
\label{theorem 2}
For the cache-aided MuPIR problem with $K=2$ messages, $K_{\rm u}=2$ users and $N\ge 2$ databases, the optimal memory-load trade-off under uncoded cache placement is characterized as
\begin{equation}\label{eqn: achiavable rate uncoded cache}
R^{\star}_{\textrm{uncoded}}(M) =
    \begin{cases}
      2 -\frac{3}{2}M ,& 0\le M\le \frac{2(N-1)}{2N-1}  \\
      \frac{(N+1)(2-M)}{2N}, & \frac{2(N-1)}{2N-1}\le M\le 2
    \end{cases}
\end{equation}
\end{corollary}
\begin{IEEEproof}
For achievability, the corner points in Theorem~\ref{theorem 2} are the memory-load pairs $(0,2)$, $\left(\frac{2(N-1)}{2N-1},\frac{N+1}{2N-1}\right)$, and $(2,0)$, which can be achieved by the same scheme as Theorem~\ref{theorem 1}.  It can be seen in Section~\ref{Section: achievability} that the achievable schemes for these corner points
are uncoded, and by memory sharing, the load of Theorem \ref{theorem 2} can be achieved.  %(See  Fig.~\ref{fig: load 1} for a detailed illustration of the load curve).

{Under the assumption of uncoded cache placement,  as shown in~\eqref{eqn: existing uncoded converse}, the converse bound  for the
shared-link coded caching problem without privacy constraint in~\cite{wan2016optimality,8226776} is also a converse for our considered caceh-aided MuPIR problem.}  When $0\le M \le 1$, it was proved in~\cite{wan2016optimality,8226776} that
\begin{align}
R^{\star}_{\textrm{uncoded}}(M) \geq    2 -\frac{3}{2} M.\label{eq:uncoded first seg}
\end{align}
In addition, the converse bound for the single-user cache-aided PIR problem  in \cite{tandon2017capacity} is also
a converse for the considered problem since increasing the number of users
cannot reduce the load.  When $0\le M\le 2$,
\begin{align}
R^{\star}_{\textrm{uncoded}}(M) \geq \left(1-\frac{M}{2}\right)\left( 1+\frac{1}{N}  \right)=\frac{(N+1)(2-M)}{2N}.\label{eq:uncoded sec seg}
\end{align}
By combining
\eqref{eq:uncoded first seg} and~\eqref{eq:uncoded sec seg},
\eqref{eqn: achiavable rate uncoded cache} can be obtained, which coincides with the achievability.
\end{IEEEproof}

For general parameter settings, we propose a {\em Product Design (PD)} design (Section~\ref{sec: theorem 4}). The corresponding achievable load is given by the following theorem.
\begin{theorem}
\label{theorem 4}
The proposed product design achieves the load of $R_{\rm PD}(M)=\min\{K(1-\frac{M}{K}),\widehat{R}(M)\}$ in which
\be \label{eqn: product design load}
\widehat{R}(M)=\frac{K_{\rm u}-t}{t+1} \left( 1+\frac{1}{N}+\cdots+\frac{1}{N^{K-1}}\right ), \ee
where $t=\frac{K_{\rm u}M}{K}\in\mathbb{Z}^+$. For non-integer values of $t$, the lower convex envelope the integer points $\left(\frac{tK}{K_{\rm u}}, \widehat{R}(M)\right)$ in which $\frac{tK}{K_{\rm u}}\in[K_{\rm u}]$ can be achieved. Moreover,
$\frac{R_{\rm PD}(M)}{R^{\star}(M)}\le 8$.
\end{theorem}
\begin{IEEEproof}
 \paragraph*{Achievability}
See Section~\ref{sec: theorem 4} for the achievable scheme to achieve $\widehat{R}(M)$.
A naive scheme suffices to achieve the load $K(1-\frac{M}{K})$ which is described as follows. Let each user cache the same $\frac{M}{K}$ portion of each message. In the private delivery phase, the remaining $1-\frac{M}{K}$ portion of each message is broadcasted to the users. It can be
seen that each user can correctly decode all $K$ messages. Hence, the scheme is private and the achievable load is $K(1-\frac{M}{K})$.

{\paragraph*{Converse}
 We use the converse bound in \cite{ghasemi2017improved} on the worst-case load for coded caching without privacy constraint, denoted by $R_{\rm caching}(M)$.
As shown in~\eqref{eqn: existing converse}, $R_{\rm caching}(M)$ is also a converse bound for the considered MuPIR problem.
  In addition, it was proved in \cite{ghasemi2017improved} that  $  R_{\rm caching}(M)$ is no less than the lower convex envelop of $\frac{1}{4}\min \{\frac{K_{\rm u}-t}{t+1},K(1-\frac{M}{K})\}$ where $t \in [K_{\rm u}]$. Hence, we have
\begin{align}
  \frac{R_{\rm PD}(M)}{R^{\star}(M)}&\le \frac{R_{\rm PD}(M)}{R_{\rm caching}(M)} \le 4\cdot \left( 1+\frac{1}{N}+\cdots+\frac{1}{N^{K-1}}\right )  \leq 8, \label{eq:use individual theta3}
\end{align}
which comes from that $R_{\rm PD}(M)=\min\{K(1-\frac{M}{K}),\widehat{R}(M)\}$,
and that $1+\frac{1}{N}+\cdots+\frac{1}{N^{K-1}}\le 2$ when $N\ge 2$.
}
\end{IEEEproof}

%-------------------------------------------------
\begin{corollary}
\label{corollary: product design opt in high cache regime}
\emph{The proposed product design is optimal
when $\frac{(K_{\rm u}-1)}{K_{\rm u}}K\le M\le K$.}
\end{corollary}
\begin{IEEEproof}
When $K\ge K_{\rm u}$ and $M=\frac{K(K_{\rm u}-1)}{K_{\rm u}}$,
(\ref{eqn: product design load}) becomes $\widehat{R}(M)= \frac{1}{K_{\rm u}}\left( 1+\frac{1}{N}+\cdots+\frac{1}{N^{K-1}} \right)$. On the other hand, the author in \cite{tandon2017capacity} showed that when $K_{\rm u}=1$, the optimal single-user cache-aided PIR load is equal to
$
\left(1-\frac{M}{K}\right)\left( 1+\frac{1}{N}+\cdots+\frac{1}{N^{K-1}} \right)$,
which also equals to (\ref{eqn: product design load}) when $M=\frac{K(K_{\rm u}-1)}{K_{\rm u}}$.
By memory sharing between $(\frac{K(K_{\rm u}-1)}{K_{\rm u}},\widehat{R}(\frac{K(K_{\rm u}-1)}{K_{\rm u}}))$ and $(K,0)$, we conclude that the proposed PD is optimal when $\frac{(K_{\rm u}-1)}{K_{\rm u}}K\le M\le K$.
\end{IEEEproof}

\paragraph*{Numerical Evaluations}
In Fig.~\ref{fig: load 1}, we consider   the MuPIR systems with $K = K_u=2$,  where   $N=2$ in Fig.~\ref{fig: load_2DBs} and $N=3$ in Fig.~\ref{fig: load_3DBs} respectively.
We compare the proposed CIA based scheme in Theorem~\ref{theorem 1}, the   optimal scheme under uncoded cache placement in Corollary~\ref{eqn: achiavable rate uncoded cache}, the product design in Theorem~\ref{theorem 4}, and the   computer-aided converse in \cite{tph:19:cai}. In addition,
it will be clarified in Theorem~\ref{theorem 3} that for the case  $K = K_u=N=2$, if  the users demand distinct messages, the optimal memory-load trade-off is  $  R_{\rm d}^{\star}(M) $ given in~\eqref{eqn: optimal distinct demands}.

  In Fig.~\ref{fig: load_2DBs}, there are two non-trivial corner points $(\frac{1}{4},\frac{3}{2})$ and $(\frac{2}{3},1)$ associated with the CIA based scheme in Theorem~\ref{theorem 1}. It can be seen that in the case of general demands, the CIA based scheme outperforms both    the optimal scheme under uncoded cache placement in Corollary~\ref{eqn: achiavable rate uncoded cache} and  the product design in Theorem~\ref{theorem 4}.  When $\frac{1}{4}\le M \le 2$, $R_{\rm CIA}$ coincides with the computer-aided converse \cite{tph:19:cai} and hence is optimal. It also can be seen that a lower load can be achieved when users only have distinct demands.
Fig.~\ref{fig: load_3DBs} shows the case when $N=3$ in which $R_{\rm CIA}$ is optimal when $\frac{1}{3}\le M\le 2$ by the computer-aided converse.
\begin{figure}
\begin{subfigure}{.5\textwidth}
  \centering
  \includegraphics[width=8cm]{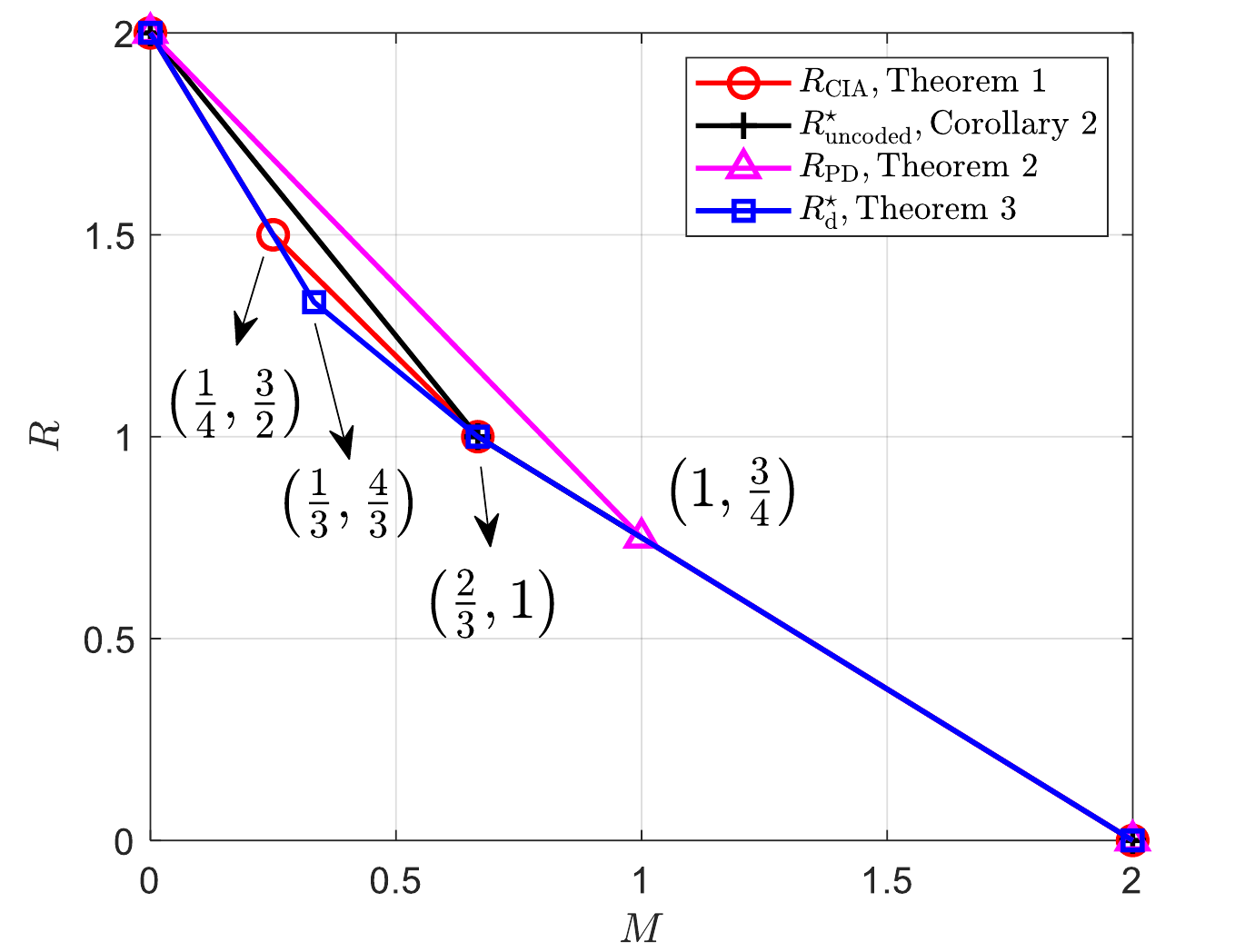}
  \vspace{-0.4cm}
  \caption{$K=K_{\rm u}=2,N=2$.}
\label{fig: load_2DBs}
\end{subfigure}
\begin{subfigure}{.5\textwidth}
  \centering
  \includegraphics[width=8cm]{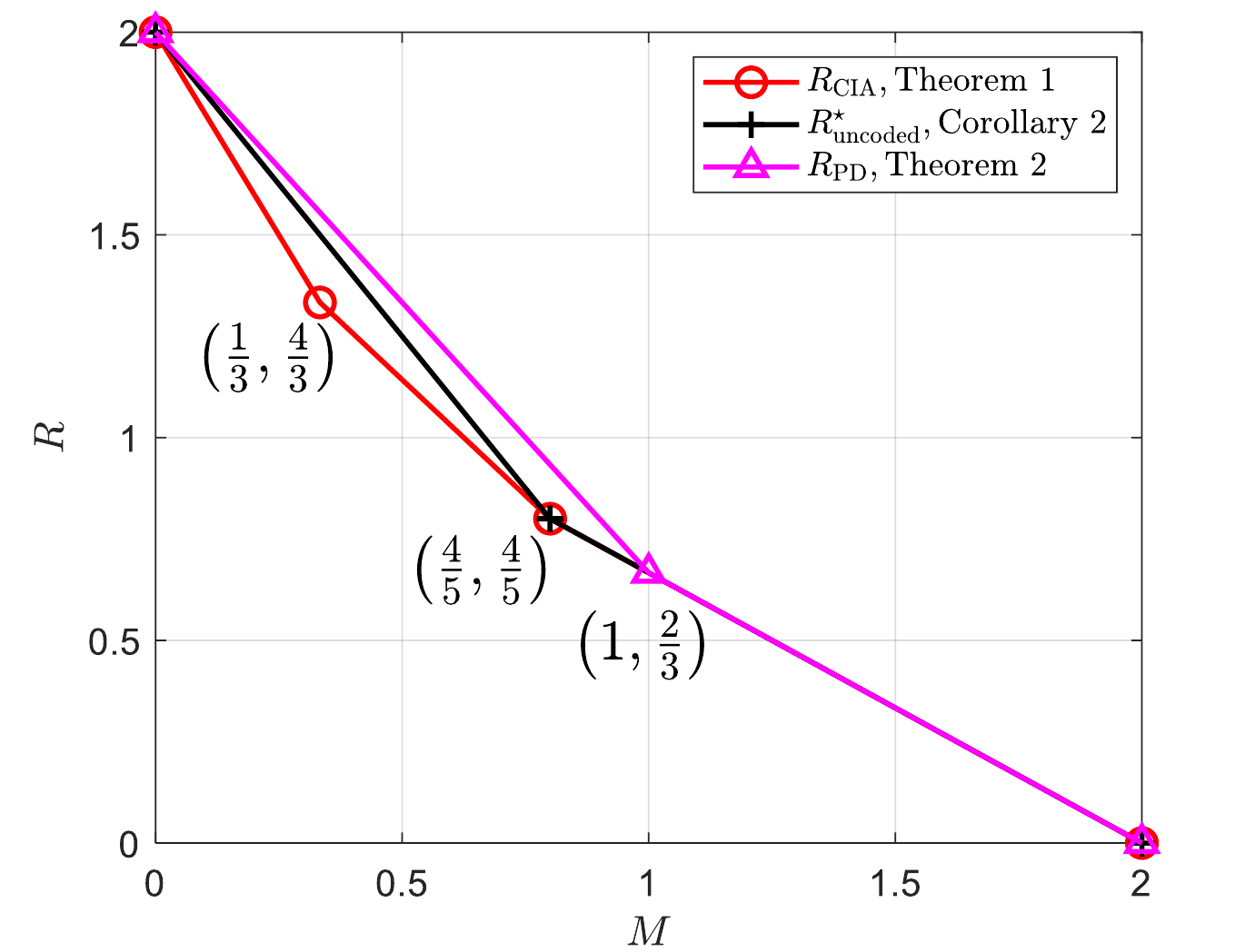}
  \vspace{-0.4cm}
  \caption{$K=K_{\rm u}=2,N=3$.}
\label{fig: load_3DBs}
\end{subfigure}
\vspace*{-0.5cm}
\caption{~\small An illustration of the achievable load $R$ of the MuPIR system with $K = K_u=2$. (a) $N=2$. Both the CIA scheme ($R_{\rm CIA}$) and the the optimal scheme under distinct demands  ($R_{\rm d}^{\star}(M) $) have four corner pints; Both the optimal scheme under uncoded cache placement ($R_{\textrm{uncoded}}^{\star}$) and the product design ($R_{\rm PD}$) have three corner points;  (b) $N=3$. $R_{\rm CIA}$ has four corner points. $R_{\rm uncoded}^{\star}$ and $R_{\rm PD}$ have three corner points.}
\label{fig: load 1}
\vspace{-0.75cm}
\end{figure}

In Fig.~\ref{fig: load_PD}, we compare the load of the product design with the best-known caching bound provided in \cite{yu2018characterizing} when $K=K_{\rm u}=6, N=2$. More specifically, Theorem 2 of \cite{yu2018characterizing} gives the caching converse as a lower convex envelope of the set of memory-load pairs
\be \left\{ \left( \frac{7-\ell}{s},\frac{s-1}{2}+\frac{\ell (\ell -1)}{2s}\right)    \Big |\forall s\in[1:6],\forall \ell\in[1:s]  \right\}\cup \left\{(0,6)\right\}.\ee Since $R_{\rm PD}(1)$ lies below the line segment connecting the memory-load pairs $(0,6)$ and $(2, R_{\rm PD}(2))$, we can use memory sharing to achieve a better load for $M=1$.

\begin{figure}
  \centering
  \includegraphics[width=8cm]{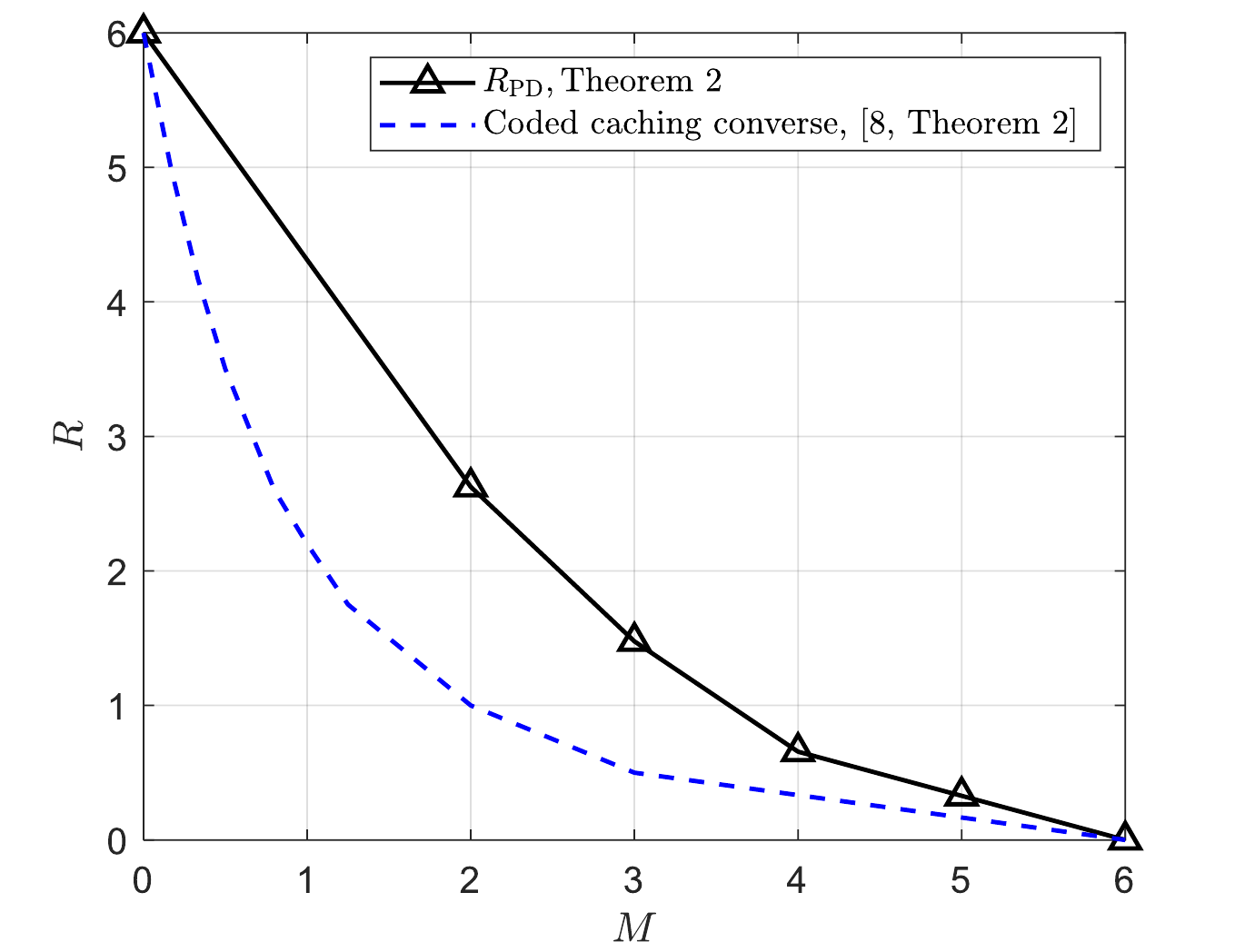}
  \vspace{-0.4cm}
  \caption{~\small The  load of  the product design compared to the best-known caching bound \cite{yu2018characterizing}  for $K=K_{\rm u}=6, N=2$.}
\label{fig: load_PD}
\vspace{-0.75cm}
\end{figure}

%%==================================================================================================
\section{Proof of Theorem \ref{theorem 1}: Description of the CIA scheme}
\label{Section: achievability}
In this section, we
prove Theorem \ref{theorem 1}. For the cache-aided MuPIR problem with $K=2$ messages, $K_{\rm u}=2$ users and $N\ge 2$ DBs, we first show the achievability of the memory-load pairs $\left(\frac{N-1}{2N},\frac{N+1}{N}\right)$ and $\left(\frac{2(N-1)}{2N-1},\frac{N+1}{2N-1}\right)$ using the proposed CIA scheme.
Note that when $M=0$, we let any one of the DBs broadcast the two messages to the users, so the memory-load pair $(0,2)$ is achievable.
When $M=2$, we let both users store the two messages in the placement phase and there is no need for the DBs to transmit anything.
Therefore, $(2,0)$ is achievable. By memory sharing between the above four corner points, the load of Theorem \ref{theorem 1} can be achieved. For each of the above two non-trivial corner points, we first describe the general achievable schemes for arbitrary number of DBs and then present an example to highlight the design intuition. Computer-aided investigation \cite{tph:19:cai} shows that the achievable load in Theorem \ref{theorem 1} is optimal when $N=2$ and 3. For general values of $N$, the converse remains open.

%---------------------------------------
\subsection{Achievability of $\left(\frac{N-1}{2N},\frac{N+1}{N}\right)$  }
%----------- ------ exmaple 1----------------------
\if{0}
We first consider the case of $N=2$, i.e., $\left(\frac{N-1}{2N},\frac{N+1}{N}\right)=(\frac{1}{4},\frac{3}{2}
)$, and then proceed to the case of arbitrary $N\ge 2$.

\begin{example}\label{example: N=2, (1/4,3/2)} (Achievability of $(\frac{1}{4},\frac{3}{2})$ for $N=2$)
Consider the cache-aided MuPIR problem with $K=2$ messages, $K_{\rm u}=2$ users and $N=2$ DBs. Let $W_1=A$ and $W_2=B$ denote the two messages and $\bm{\theta}=(\theta_1,\theta_2)\in[2]^2$ be the user demands.
The cache placement and private delivery phases using CIA are described as follows.

\emph{1) Cache placement:} Assume that each message consists of $L=4$ bits, i.e., $A=(A_1,A_2,A_3,A_4)$, $B=(B_1,B_2,B_3,B_4)$. Each user stores a linear combination of the message bits, i.e.,
$Z_1 = \{A_1+ A_2+B_1+B_2 \}, Z_2 = \{A_3+ A_4+B_3+B_4 \}$.
%\end{eqnarray}
It can be seen that the cache memory constraint $M=\frac{1}{4}$ is satisfied.

\emph{2) Private delivery:} In this phase, the users download an answer (coded bits) from each DB.
In particular, the answer of DB 1 consists of two random linear combinations, i.e., $A_1^{[\bm{\theta}]}=(A_{1,1}^{[\bm{\theta}]}, A_{1,2}^{[\bm{\theta}]})$, where
$A_{1,1}^{[\bm{\theta}]} = \uv_{1,1}A_{(1:2)}^{\rm T} +  \vv_{1,1}{B}_{(1:2)}^{\rm T},
A_{1,2}^{[\bm{\theta}]} = \uv_{1,2}A_{(3:4)}^{\rm T} +  \vv_{1,2}{B}_{(3:4)}^{\rm T}$.
The linear combination coefficients $\uv_{1,1},\vv_{1,1,}\uv_{1,2},\vv_{1,2}\in\mathbb{F}_2^{1\times 2}\backslash \{[0,0]\}$ are subject to design according to different user demands.
The answer of DB 2 consists of four linear combinations, i.e.,  $A_2^{[\bm{\theta}]}=(A_{2,1}^{[\bm{\theta}]},A_{2,2}^{[\bm{\theta}]},A_{2,3}^{[\bm{\theta}]},A_{2,4}^{[\bm{\theta}]})$  which are
%\begin{eqnarray}
$A_{2,1}^{[\bm{\theta}]} = \gv_1A_{(1:2)}^{\rm T},
A_{2,2}^{[\bm{\theta}]} = \gv_2B_{(1:2)}^{\rm T},
A_{2,3}^{[\bm{\theta}]} = \gv_3A_{(3:4)}^{\rm T},
A_{2,4}^{[\bm{\theta}]} = \gv_4B_{(3:4)}^{\rm T}$,
%\end{eqnarray}
in which the coefficients $\gv_j\in\mathbb{F}_2^{1\times 2}\backslash \{[0,0]\},\forall j\in[4]$ are subject to design.
The answers can be written in a more compact form as
\begin{equation}\label{Eq: answer matrix (1/4,3/2)}
{\small
\begin{bmatrix}
A_{1,1}^{[\bm{\theta}]}\\
A_{1,2}^{[\bm{\theta}]}\\
A_{2,1}^{[\bm{\theta}]}\\
A_{2,2}^{[\bm{\theta}]}\\
A_{2,3}^{[\bm{\theta}]}\\
A_{2,4}^{[\bm{\theta}]}
\end{bmatrix}=\begin{bmatrix}
\uv_{1,1}  & \bm{0}_2 & \vv_{1,1}  & \bm{0}_2\\
 \bm{0}_2 & \uv_{1,2} & \bm{0}_2  & \vv_{1,2}\\
\gv_1  & \bm{0}_2 &  \bm{0}_2  & \bm{0}_2\\
\bm{0}_2  & \bm{0}_2 & \gv_2  & \bm{0}_2\\
\bm{0}_2  & \gv_3 &\bm{0}_2 & \bm{0}_2\\
\bm{0}_2  & \bm{0}_2 & \bm{0}_2  &\gv_4\\
\end{bmatrix}\begin{bmatrix}
A_{(1:4)}^{\rm T}\\
B_{(1:4)}^{\rm T}
\end{bmatrix}.
}
\end{equation}
We next show how these random linear coefficients can be designed such that both users can correctly decode their desired messages. We defer the description of the privacy until we complete the the decodability argument for different user demands.

Assume $(\theta_1,\theta_2)=(1,2)$, i.e., users 1 and 2 request $A$ and $B$ respectively. To guarantee decodabilty and privacy, two sets of constraints, i.e., the full-rank condition and alignment condition should be satisfied. More specifically, the full-rank condition requires that the following six coefficient matrices to have full rank:\footnote{{ The way we choose the coefficients to satisfy these requirements will be described later. Here we only present the conditions which are necessary for correct decoding and preserving demand privacy.}}
\be\label{Eq: full rank [1,2], (1/4,3/2)}
\textrm{\emph{Full-rank condition}:} \quad
\begin{bmatrix}
[1,1]\\
\gv_1
\end{bmatrix} ,
\begin{bmatrix}
[1,1]\\
\gv_4
\end{bmatrix},
\begin{bmatrix}
\uv_{1,2}\\
\gv_3
\end{bmatrix},
\begin{bmatrix}
\vv_{1,1}\\
\gv_2
\end{bmatrix},
{\begin{bmatrix}
[1,1]\\
\gv_2
\end{bmatrix},
\begin{bmatrix}
[1,1]\\
\gv_3
\end{bmatrix}}.
\ee
The coefficients $\uv_{1,1}$ and $\vv_{1,2}$ are chosen as
\be \label{Eq: alignment [1,2], (1/4,3/2)}
\textrm{\emph{Alignment condition}:} \quad \uv_{1,1}=\gv_1 ,\;\vv_{1,2}=\gv_4 .
\ee
Next we show that both users can correctly decode their desired messages using the above two conditions.

From (\ref{Eq: answer matrix (1/4,3/2)}), we obtain
\be
\begin{bmatrix}\label{Eq: theta=(1,2), (1/4,3/2), solve A(3,4)}
A_{1,2}^{{[(1,2)]}}\\
A_{2,3}^{{[(1,2)]}}
\end{bmatrix}=\begin{bmatrix}
\uv_{1,2} & \vv_{1,2}\\
 \gv_3  & \bm{0}_2
\end{bmatrix}\begin{bmatrix}
A_{(3:4)}^{\rm T}\\
B_{(3:4)}^{\rm T}
\end{bmatrix} .
\ee
Since $\gv_4=\vv_{1,2}$, we have $A_{2,4}^{[(1,2)]}=\gv_4B_{(3:4)}^{\rm T}=\vv_{1,2}B_{(3:4)}^{\rm T}$. Subtracting $A_{2,4}^{[(1,2)]}$ from $A_{1,2}^{{[(1,2)]}}$, and due to $\textrm{rank}([\uv_{1,2};\gv_3])=2$, the two messages bits $A_3$ and $A_4$ can be solved from
(\ref{Eq: theta=(1,2), (1/4,3/2), solve A(3,4)}) as
\be
\begin{bmatrix}
A_3\\
A_4
\end{bmatrix}=
\begin{bmatrix}
\uv_{1,2} \\
 \gv_3
\end{bmatrix}^{-1}
\begin{bmatrix}
A_{1,2}^{[(1,2)]}-A_{2,4}^{[(1,2)]}\\
A_{2,3}^{[(1,2)]}
\end{bmatrix}.
\ee
Similarly, due to $\gv_1=\uv_{1,1}$, we have $A_{2,1}^{[(1,2)]}=\gv_1A_{(1:2)}^{\rm T}=\uv_{1,1}A_{(1:2)}^{\rm T}$. Subtracting $A_{2,1}^{[(1,2)]}$ from $A_{1,1}^{[(1,2)]}$ and due to ${\rm rank}([\vv_{1,1};\gv_2])=2$, $B_1$ and $B_2$ can be decoded as
\be
\begin{bmatrix}
B_1\\
B_2
\end{bmatrix}=
\begin{bmatrix}
\vv_{1,1} \\
 \gv_2
\end{bmatrix}^{-1}
\begin{bmatrix}
A_{1,1}^{[(1,2)]}-A_{2,1}^{[(1,2)]}\\
A_{2,2}^{[(1,2)]}
\end{bmatrix}.
\ee
Therefore, both users can correctly decode $A_3,A_4$ and $B_1,B_2$.

Now, user 1 still needs $A_1$ and $A_2$ while user 2 needs $B_3$ and $B_4$. For user 1, the interference $B_1+B_2$ can be removed from $Z_1$ and obtain
$A_1+A_2$.
Together with $A_{2,1}^{[(1,2)]}$ and due to ${\rm rank}([[1,1];\gv_1])=2$, user 1 can decode $A_1,A_2$ from\footnote{{ With a slight abuse of notation, in
(\ref{eqn: decoded A(1:2) for user 1}), we used $Z_1$ to represent the linear combination stored by user 1. %instead being a set.
}}
\be \label{eqn: decoded A(1:2) for user 1}
\begin{bmatrix}
A_1\\
A_2
\end{bmatrix}
= \begin{bmatrix}
[1,1]\\
\gv_1
\end{bmatrix}^{-1}
\begin{bmatrix}
Z_1-(B_1+B_2)\\
A_{2,1}^{[(1,2)]}
\end{bmatrix}.
\ee
Similarly, for user 2, the interference $A_3+A_4$ can be removed from $Z_2$ since they are already available. Together with $A_{2,4}^{[(1,2)]}$ and due to ${\rm rank}([[1,1];\gv_4])=2$, user 2 can decode $B_3,B_4$ from
\be \label{eqn: decoded B(3:4) for user 2}
\begin{bmatrix}
B_3\\
B_4
\end{bmatrix}
= \begin{bmatrix}
\bm{\beta}_2\\
\gv_4
\end{bmatrix}^{-1}
\begin{bmatrix}
Z_2-(A_3+A_4)\\
A_{2,4}^{[(1,2)]}
\end{bmatrix}.
\ee
Therefore, the users can correctly recover their desired messages.\footnote{{ Note that for the decoding of $\bm{\theta}=%(\theta_1,\theta_2)
(1,2)$, we only used the full-rank property of the first four coefficient matrices in the full-rank condition
(\ref{Eq: full rank [1,2], (1/4,3/2)}). The reason that we require the two additional matrices to be full-rank is that these two matrices must be full-rank in order for the correct decoding of other demands. If they are not full-rank, then DB 2 can exclude the possibilities of certain user demands, rendering the scheme non-private.}} Note that by far we have only imposed certain conditions on the linear coefficients to enforce correct decoding and the exact values of the coefficients have not been specified yet, which will be described later.

For all other demands, the private delivery phase can be designed similarly.

\if{0}
For $(\theta_1,\theta_2)=(2,1)$, the following six matrices
\be\label{Eq: full rank [2,1], (1/4,3/2)}
\begin{bmatrix}
[1,1]\\
\gv_3
\end{bmatrix} ,
\begin{bmatrix}
[1,1]\\
\gv_2
\end{bmatrix},
\begin{bmatrix}
\uv_{1,1}\\
\gv_1
\end{bmatrix},
\begin{bmatrix}
\vv_{1,2}\\
\gv_4
\end{bmatrix},
{\begin{bmatrix}
[1,1]\\
\gv_1
\end{bmatrix},
\begin{bmatrix}
[1,1]\\
\gv_4
\end{bmatrix}}.
\ee
are required to be full-rank. The coefficients $\uv_{1,2}$ and $\vv_{1,1}$ are chosen as
\be \label{Eq: alignment [2,1], (1/4,3/2)}
\uv_{1,2}=\gv_3 ,\;\vv_{1,1}=\gv_2 .
\ee
With such an alignment, similar to the case of $(\theta_1,\theta_2)=(1,2)$, it can be easily verified that the two users can correctly decode the messages $B$ and $A$ respectively.

For $(\theta_1,\theta_2)=(1,1)$, the following six matrices
\be\label{Eq: full rank [1,1], (1/4,3/2)}
\begin{bmatrix}
\uv_{1,1}\\
\gv_1
\end{bmatrix} ,
\begin{bmatrix}
\uv_{1,2}\\
\gv_3
\end{bmatrix},
\begin{bmatrix}
[1,1]\\
\gv_1
\end{bmatrix},
\begin{bmatrix}
[1,1]\\
\gv_4
\end{bmatrix},
{\begin{bmatrix}
[1,1]\\
\gv_2
\end{bmatrix},
\begin{bmatrix}
[1,1]\\
\gv_3
\end{bmatrix}}.
\ee
are required to be full-rank and the coefficients $\vv_{1,1}$ and $\vv_{1,2}$ are chosen as
\be \label{Eq: alignment [1,1], (1/4,3/2)}
 \vv_{1,1}=\gv_2 ,\;\vv_{1,2}=\gv_4 .
\ee
Similarly, it can be easily verified that both users can correctly decode message $A$.

For $(\theta_1,\theta_2)=(2,2)$, the following six matrices
\be\label{Eq: full rank [2,2], (1/4,3/2)}
\begin{bmatrix}
\vv_{1,1}\\
\gv_2
\end{bmatrix} ,
\begin{bmatrix}
\vv_{1,2}\\
\gv_4
\end{bmatrix},
\begin{bmatrix}
[1,1]\\
\gv_1
\end{bmatrix},
\begin{bmatrix}
[1,1]\\
\gv_4
\end{bmatrix},
{\begin{bmatrix}
[1,1]\\
\gv_2
\end{bmatrix},
\begin{bmatrix}
[1,1]\\
\gv_3
\end{bmatrix}}.
\ee
are required to be full-rank and the coefficients $\uv_{1,1}$ and $\uv_{1,2}$ are chosen as
\be \label{Eq: alignment [2,2], (1/4,3/2)}
 \uv_{1,1}=\gv_1 ,\;\uv_{1,2}=\gv_3 .
\ee
It can be easily verified that both users can correctly decode message $B$.
\fi

Due to the explicit constraints on the structure of the linear coefficients for correct decoding corresponding to different user demands, we have the following observation.

\begin{observation}\emph{
Due to the full-rank conditions (e.g.,
(\ref{Eq: full rank [1,2], (1/4,3/2)}), the DB linear coefficients must satisfy that $\uv_{1,i},\vv_{1,i}\neq [1,1],[0,0],\forall i\in[2]$ and $\gv_j\neq [1,1],[0,0],\forall j\in[4]$, i.e., the linear coefficient vectors can only take values $[0,1]$ or $[1,0]$.
Therefore, there are $2^4=16$ different choices for choosing DB 1 coefficients $(\uv_{1,1},\vv_{1,1},\uv_{1,2},\vv_{1,2})\in\{[0,1],[1,0]\}^4$, and also $2^4=16$ choices for choosing DB 2 coefficients $(\gv_1,\gv_2,\gv_3,\gv_4)\in \{[0,1],[1,0]\}^4$. On   one hand, for each DB 1 choice, the DB 2 linear coefficients $\gv_j,\forall j\in[4]$ can be uniquely determined through the full-rank and alignment conditions corresponding to each $(\theta_1,\theta_2)$. For example,  consider a specific choice of the DB 1 linear coefficients $\uv_{1,1}=\uv_{1,2}=[0,1]$, $\vv_{1,1}=\vv_{1,2}=[1,0]$. For $(\theta_1,\theta_2)=(1,2)$, by the alignment condition
(\ref{Eq: alignment [1,2], (1/4,3/2)}), we can immediately obtain $\gv_1=\uv_{1,1}=[0,1]$, $\gv_4=\vv_{1,2}=[1,0]$. Since the matrices $[\uv_{1,2};\gv_3]$ and $[\vv_{1,1};\gv_2]$ have full ranks, we have $\gv_2=[0,1],\gv_3=[1,0]$. Similarly, for $(\theta_1,\theta_2)=(2,1)$, we have $\gv_1=\gv_2=[1,0]$, $\gv_3=\gv_4=[0,1]$. For $(\theta_1,\theta_2)=(1,1)$, we have $\gv_1=\gv_2=\gv_3=\gv_4=[1,0]$. For $(\theta_1,\theta_2)=(2,2)$, we have $\gv_1=\gv_2=\gv_3=\gv_4=[0,1]$. On the other hand, for each DB 2 choice, the DB 1 linear coefficients $\uv_{1,i},\vv_{1,i},\forall i\in[2]$ can be uniquely determined through the full-rank and alignment conditions corresponding to each $(\theta_1,\theta_2)$}. \end{observation}

With the above observation, we
can specify the choice of the linear coefficients. We first define two matrices $\mathbf{D}_1,\mathbf{D}_2\in\mathbb{F}^{4\times 8}_2$ as follows:
\be
\label{Eq: candidate matrix, (1/4,3/2)}
{\small
\mathbf{D}_1\triangleq
\begin{bmatrix}
 0 & 1 & 0 & 1 & 1 & 0& 1& 0\\
  1 & 0 & 1 & 0 & 0 & 1& 0& 1\\
   1 & 0 & 0 & 1 & 0 & 1& 1& 0\\
    0 & 1 & 1 & 0 & 1 & 0& 0& 1
\end{bmatrix},\;
\mathbf{D}_2\triangleq
\begin{bmatrix}
 0 & 1 & 0 & 1 & 1 & 0& 1& 0\\
  1 & 0 & 1 & 0 & 0 & 1& 0& 1\\
   1 & 0 & 1 & 0 & 1 & 0& 1& 0\\
    0 & 1 & 0 & 1 & 0 & 1& 0& 1
\end{bmatrix}.
}
\ee
The two matrices correspond to the possible choices of linear coefficients of the two DBs respectively. More specifically, each row of $\mathbf{D}_1$ represents a different realization of the DB 1 coefficient vector $[\uv_{1,1},\uv_{1,2},\vv_{1,1},\vv_{1,2}]$ and each row of $\mathbf{D}_2$ represents a realization of the DB 2 coefficient vector $[\gv_1,\gv_2,\gv_3,\gv_4]$.
Let the DB 1 coefficient vector $[\uv_{1,1},\uv_{1,2},\vv_{1,1},\vv_{1,2}]$ be randomly chosen from the rows of $\mathbf{D}_1$ with equal probabilities. Depending on the DB 1 choice and user demands, the DB 2 coefficient vector $[\gv_1,\gv_2,\gv_3,\gv_4]$ can be determined. For example,
when $(\theta_1,\theta_2)=(1,2)$, if $\mathbf{D}_1(1,:)$ is chosen, let $[\gv_1,\gv_2,\gv_3,\gv_4]=\mathbf{D}_2(1,:)$.

We next prove the correctness and privacy the above delivery scheme.

\emph{Correctness:} Decodabilility is straightforward since the construction of the linear coefficients in $\mathbf{D}_1$ and $\mathbf{D}_2$ follows the full-rank and alignment conditions for any user demands $(\theta_1,\theta_2)\in[2]^2$.

\emph{Privacy:} First, the above delivery phase is private from the perspective of DB 1. The reason is as follows. For any choice of the answer coefficient vector of DB 1, there exists four different configurations of the  DB 2 answer coefficient vector each of which corresponds to a different user demand vector $(\theta_1,\theta_2)\in[2]^2$. Since DB 1 is unaware of the choice of DB 2, it does not know which
demands $(\theta_1, \theta_2)$ are being requested. Since $[\uv_{1,1},\uv_{1,2},\vv_{1,1},\vv_{1,2}]$ is chosen randomly (with equal probabilities) from the rows of $\mathbf{D}_1$, and the user demand vector $(\theta_1,\theta_2)$ is assumed to be generated randomly and uniformly from $[2]^2$, the actual user demands $(\theta_1,\theta_2)$ is equally likely to be $(1,2),(2,1),(1,1)$ or $(2,2)$ from DB 1's perspective. Following a similar argument, we can see that the delivery phase is also private from DB 2's perspective. Therefore, the above delivery phase is private.

\emph{Performance:} Since $D=6$ linear combinations are downloaded in total, the achieved load is $R=\frac{D}{L}=\frac{3}{2}$.
\hfill  $\lozenge $
\end{example}
\fi

%----------------------------------------------------------------------------------------------------------------------------
Let $W_1=A$ and $W_2=B$ denote the two messages, each consisting of
$L=2N$ bits, i.e.,
$A=(A_1,A_2,\cdots, A_{2N}),
B=(B_1,B_2,\cdots, B_{2N})$.
The proposed scheme is described as follows.

\emph{1) Cache placement:} Each user stores $N-1$ liner combinations of the message bits in its cache (therefore $M=\frac{N-1}{2N}$), i.e.,
\be Z_1 =\left\{\bm{\alpha}_{1,j}A_{(1:N)}^{\rm T} + \bm{\beta}_{1,j}B_{(1:N)}^{\rm T} : j\in[N-1]\right \}, \ee
and
 \be Z_2 = \left \{\bm{\alpha}_{2,j}A_{(N+1:2N)}^{\rm T} + \bm{\beta}_{2,j}B_{(N+1:2N)}^{\rm T} : j\in[N-1]\right \},\ee
in which the linear combination coefficients $\bm{\alpha}_{i,j},\bm{\beta}_{i,j}\in\mathbb{F}_2^{1\times N}\backslash \{\mathbf{0}_N\},\forall i\in[2],\forall j\in[N-1]$ are chosen such that $
\textrm{rank}([\bm{\alpha}_{i,1};\bm{\alpha}_{i,2};\cdots;\bm{\alpha}_{i,N-1}])=N-1 $ and $\textrm{rank}([\bm{\beta}_{i,1};\bm{\beta}_{i,2};\cdots;\bm{\beta}_{i,N-1}])=N-1,\forall i\in[2]$. WLOG, $\forall i=1,2$, we choose the cache coefficients to be
$
[\bm{\alpha}_{i,1};
\bm{\alpha}_{i,2};
\cdots;
\bm{\alpha}_{i,N-1}]
 =
[\bm{\beta}_{i,1};
\bm{\beta}_{i,2};
\cdots;
\bm{\beta}_{i,N-1}]
=
[\mathbf{I}_{N-1}, \bm{0}_{N-1}^{\rm T}]
$.
Furthermore, let $Z_{i,1}$, $Z_{i,2}$, $\cdots$, $Z_{i,N-1}$ denote the $N-1 $
linear combinations in $Z_i,\forall i\in[2]$, i.e., $\forall j\in[N-1]$, we have $Z_{1,j} =\bm{\alpha}_{1,j}A_{(1:N)}^{\rm T} + \bm{\beta}_{1,j}B_{(1:N)}^{\rm T}$ and $Z_{2,j} =\bm{\alpha}_{2,j}A_{(N+1:2N)}^{\rm T} + \bm{\beta}_{2,j}B_{(N+1:2N)}^{\rm T}$.

\emph{2) Private delivery:} In this phase, the two users download an answer from each DB according to their demands $(\theta_1,\theta_2)$. The answers are random linear combinations of certain message bits.
In particular, the answer of DB $n\in[N-1]$ consists of two random linear combinations, i.e., $A_n^{[\bm{\theta}]}\eqdef (A_{n,1}^{[\bm{\theta}]}, A_{n,2}^{[\bm{\theta}]})$ in which
$A_{n,1}^{[\bm{\theta}]} =\uv_{n,1}A_{(1:N)}^{\rm T}+ \vv_{n,1}B_{(1:N)}^{\rm T},A_{n,2}^{[\bm{\theta}]} =\uv_{n,2}A_{(N+1:2N)}^{\rm T}+ \vv_{n,2}B_{(N+1:2N)}^{\rm T}$.
The answer of DB $N$ consists of four random linear combinations, i.e., $A_n^{[\bm{\theta}]}\eqdef ( A_{N,1}^{[\bm{\theta}]},A_{N,2}^{[\bm{\theta}]},A_{N,3}^{[\bm{\theta}]},A_{N,4}^{[\bm{\theta}]}  )$ in which
$A_{N,1}^{[\bm{\theta}]}=\gv_1A_{(1:N)}^{\rm T},
A_{N,2}^{[\bm{\theta}]}= \gv_2B_{(1:N)}^{\rm T},
A_{N,3}^{[\bm{\theta}]}= \gv_3A_{(N+1:2N)}^{\rm T},
A_{N,4}^{[\bm{\theta}]}= \gv_4B_{(N+1:2N)}^{\rm T}$.
The linear coefficients $\gv_1,\gv_2,\gv_3,\gv_4,\uv_{n,j},\vv_{n,j}\in\mathbb{F}_2^{1\times N},\forall n\in[N-1],\forall j\in[2]$ used by the answers are subject to design according to the user demands. Therefore, in total $2N+2$ random linear combinations will be downloaded in the delivery phase. These answers can be written as
\be
\label{Eq: compact form, first point}
{\small
\begin{bmatrix}
A_{1,1}^{[\bm{\theta}]}\\
A_{1,2}^{[\bm{\theta}]}\\
\vdots\\
A_{N-1,1}^{[\bm{\theta}]}\\
A_{N-1,2}^{[\bm{\theta}]}\\
A_{N,1}^{[\bm{\theta}]}\\
A_{N,2}^{[\bm{\theta}]}\\
A_{N,3}^{[\bm{\theta}]}\\
A_{N,4}^{[\bm{\theta}]}
\end{bmatrix}=
\begin{bmatrix}
\uv_{1,1} & \mathbf{0}_{N}&\vv_{1,1}  & \mathbf{0}_{N}\\
\mathbf{0}_{N}& \uv_{1,2} & \mathbf{0}_{N} &\vv_{1,2}  \\
\vdots & \vdots &\vdots &\vdots\\
\uv_{N-1,1} & \mathbf{0}_{N}&\vv_{N-1,1}  & \mathbf{0}_{N}\\
\mathbf{0}_{N}& \uv_{N-1,2} & \mathbf{0}_{N} &\vv_{N-1,2} \\
\gv_1 & \mathbf{0}_{N}& \mathbf{0}_{N}& \mathbf{0}_{N}\\
\mathbf{0}_{N}& \mathbf{0}_{N}& \gv_2& \mathbf{0}_{N}\\
 \mathbf{0}_{N}&\gv_3 & \mathbf{0}_{N}& \mathbf{0}_{N}\\
\mathbf{0}_{N}& \mathbf{0}_{N}&  \mathbf{0}_{N}& \gv_4
\end{bmatrix}
\begin{bmatrix}
A_{(1:2N)}^{\rm T}\\
B_{(1:2N)}^{\rm T}
\end{bmatrix}.
}
\ee

We next show how the linear coefficients can be designed using the idea of CIA such that the users can correctly recover their desired messages. Due to the space limit, we will only consider $(\theta_1,\theta_2)=(1,2)$ and $(\theta_1,\theta_2)=(1,1)$. The cases of $(\theta_1,\theta_2)=(2,1)$ and $(2,2)$ follow similarly.

%-----------------------------------------------------
For $(\theta_1,\theta_2)=(1,2)$, i.e., user 1 and 2 demand messages $A$ and $B$ respectively, the following six coefficient matrices should be full-rank:
\be
\label{Eq: full rank (1,2), first point}
{\small
\textrm{\emph{Full-rank condition}:}\quad
\underbrace{\begin{bmatrix}
\bm{\alpha}_{1,1}\\
\vdots\\
\bm{\alpha}_{1,N-1}\\
\gv_1
\end{bmatrix},
 \begin{bmatrix}
\bm{\beta}_{2,1}\\
\vdots\\
\bm{\beta}_{2,N-1}\\
\gv_4
\end{bmatrix},
\begin{bmatrix}
\uv_{1,2}\\
\vdots\\
\uv_{N-1,2}\\
\gv_3
\end{bmatrix},
\begin{bmatrix}
\vv_{1,1}\\
\vdots\\
\vv_{N-1,1}\\
\gv_2
\end{bmatrix}}_{\textrm{For correct decoding of $(\theta_1,\theta_2)=(1,2)$} },
\underbrace{\begin{bmatrix}
\bm{\alpha}_{2,1}\\
\vdots\\
\bm{\alpha}_{2,N-1}\\
\gv_3
\end{bmatrix},
\begin{bmatrix}
\bm{\beta}_{1,1}\\
\vdots\\
\bm{\beta}_{1,N-1}\\
\gv_2
\end{bmatrix}}_{\textrm{For privacy}}.
}
\ee
Note that only the first four coefficient matrices being full-rank in (\ref{Eq: full rank (1,2), first point}) is mandatory to guarantee the correct decoding for the case of $(\theta_1,\theta_2)=(1,2)$. The extra two matrices $[\bm{\alpha}_{2,1};\cdots;\bm{\alpha}_{2,N-1};\gv_3]$ and $[\bm{\beta}_{1,1};\cdots;\bm{\beta}_{1,N-1};\gv_2]$ being full-rank is mandatory for the correct decoding of other user demands. The reason that we require the two extra matrices to be  full-rank for $(\theta_1,\theta_2)=(1,2)$ is that, if the two matrices are not full-rank here (the DBs can check this since we have assumed that the DBs are aware of the users' cache placement), then DB 2 can know that the actual demands being requested are $(\theta_1,\theta_2)=(1,2)$ since correct decoding is impossible for any other case of $(\theta_1,\theta_2)\ne (1,2)$. In fact, any full-rank coefficient matrix consisting of the linear coefficients of one DB and the cache coefficients which are necessary for the correct decoding of one demand vector $(\theta_1,\theta_2)$ must be full-rank for all possible demands $(\theta_1,\theta_2)\in[2]^2$ for the purpose of privacy. This multi-purpose full-rank requirement holds for all user demands. The required alignment is
\begin{subequations}
\label{Eq: alignment, (1,2), first point}
\begin{align}
\textrm{\emph{Alignment condition}:}\quad
&\gv_1 =\uv_{1,1}= \uv_{2,1}=\cdots=\uv_{N-1,1},\label{Eq: g1 alignment, (1,2), first point}\\
&\gv_4 =\vv_{1,2}= \vv_{2,2}=\cdots=\vv_{N-1,2}.\label{Eq: g4 alignment, (1,2), first point}
\end{align}
\end{subequations}
We next show that with the above full-rank and alignment conditions, the two users can correctly decode messages $A$ and $B$ respectively.

Due to the alignment condition of (\ref{Eq: g1 alignment, (1,2), first point}), we have $ A_{N,4}^{[(1,2)]}=\gv_4B_{(N+1:2N)}^{\rm T} =\vv_{1,2}B_{(N+1:2N)}^{\rm T} =\cdots=\vv_{N-1,2}B_{(N+1:2N)}^{\rm T}$, i.e., the message bits $B_{(N+1:2N)}$ are aligned among the linear combinations $A_{1,2}^{[(1,2)]},A_{2,2}^{[(1,2)]},\cdots, A_{N-1,2}^{[(1,2)]}$.
Subtracting $A_{N,4}^{[(1,2)]}$ from $A_{1,2}^{[(1,2)]},A_{2,2}^{[(1,2)]},\cdots, A_{N-1,2}^{[(1,2)]}$ in (\ref{Eq: compact form, first point}), we obtain
\be
\label{Eq: solve A(N+1:2N)}
{\begin{bmatrix}
A_{1,2}^{[(1,2)]}- A_{N,4}^{[(1,2)]}\\
\vdots\\
A_{N-1,2}^{[(1,2)]}- A_{N,4}^{[(1,2)]}\\
A_{N,3}^{[(1,2)]}
\end{bmatrix}}= {\begin{bmatrix}
\uv_{1,2}\\
\vdots\\
\uv_{N-1,2}\\
\gv_3
\end{bmatrix}}A_{(N+1:2N)}^{\rm T}.
\ee
Since the coefficient matrix on the RHS of (\ref{Eq: solve A(N+1:2N)}) is full-rank, $A_{(N+1:2N)}$ can be solved by matrix inversion as
\be
\label{Eq: A(N+1:2N) solution}
A_{(N+1:2N)}^{\rm T}={\begin{bmatrix}
\uv_{1,2}\\
\vdots\\
\uv_{N-1,2}\\
\gv_3
\end{bmatrix}}^{-1}{\begin{bmatrix}
A_{1,2}^{[(1,2)]}- A_{N,4}^{[(1,2)]}\\
\vdots\\
A_{N-1,2}^{[(1,2)]}- A_{N,4}^{[(1,2)]}\\
A_{N,3}^{[(1,2)]}
\end{bmatrix}}.
\ee
Therefore, both users can decode $A_{(N+1:2N)}$. Similarly, due to the alignment condition of (\ref{Eq: g4 alignment, (1,2), first point}), we have $A_{N,1}^{[(1,2)]}=\gv_1A_{(1:N)}^{\rm T} =\uv_{1,1}A_{(1:N)}^{\rm T} =\cdots=\uv_{N-1,1}A_{(1:N)}^{\rm T} $, i.e., the message bits $A_{(1:N)}$ are aligned among the linear combinations $A_{1,1}^{[(1,2)]},A_{2,1}^{[(1,2)]},\cdots,A_{N-1,1}^{[(1,2)]}$. Subtracting $A_{N,1}^{[(1,2)]}$ from $A_{1,1}^{[(1,2)]},A_{2,1}^{[(1,2)]},\cdots,A_{N-1,1}^{[(1,2)]}$, we obtain
\be
\label{Eq: solve B(1:N)}
{\begin{bmatrix}
A_{1,1}^{[(1,2)]}- A_{N,1}^{[(1,2)]}\\
\vdots\\
A_{N-1,1}^{[(1,2)]}- A_{N,1}^{[(1,2)]}\\
A_{N,2}^{[(1,2)]}
\end{bmatrix}}= {\begin{bmatrix}
\vv_{1,1}\\
\vdots\\
\vv_{N-1,1}\\
\gv_2
\end{bmatrix}}B_{(1:N)}^{\rm T}.
\ee
Since the linear coefficient matrix on the RHS of (\ref{Eq: solve B(1:N)}) is full-rank, $B_{(1:N)}$ can be solved as
\begin{equation}\label{eqn: B(1:N) solution}
B_{(1:N)}^{\rm T}={\begin{bmatrix}
\vv_{1,1}\\
\vdots\\
\vv_{N-1,1}\\
\gv_2
\end{bmatrix}}^{-1}{\begin{bmatrix}
A_{1,1}^{[(1,2)]}- A_{N,1}^{[(1,2)]}\\
\vdots\\
A_{N-1,1}^{[(1,2)]}- A_{N,1}^{[(1,2)]}\\
A_{N,2}^{[(1,2)]}
\end{bmatrix}}.
\end{equation}
Therefore, both users can correctly decode $B_{(1:N)}$.

Now the message bits $A_{(N+1:2N)},B_{(1:N)}$ are available to both users. User 1 still needs $A_{(1:N)}$ and user 2 still needs $B_{(N+1:2N)}$. Removing the interference of $B_{(1:N)}$ from $Z_1$, user 1 obtains $N-1$ linear combinations of $A_{(1:N)}$.
Together with $A_{N,1}^{[(1,2)]}=\gv_1A_{(1:N)}^{\rm T}$, user 1 obtains $N$ independent linear combinations of $A_{(1:N)}$, from which $A_{(1:N)}$ can be solved as
\be
\label{Eq: A(1:N) solution, first point}
A_{(1:N)}^{\rm T}=\begin{bmatrix}
\bm{\alpha}_{1,1}\\
\vdots\\
\bm{\alpha}_{1,N-1}\\
\gv_1
\end{bmatrix}^{-1} \begin{bmatrix}
Z_{1,1}-\bm{\beta}_{1,1}B_{(1:N)}^{\rm T}\\
\vdots\\
Z_{1,N-1}-\bm{\beta}_{1,N-1}B_{(1:N)}^{\rm T}\\
A_{N,1}^{[(1,2)]}\end{bmatrix}.
\end{equation}
Therefore, user 1 can correctly decode all the $2N$ bits of the desired message $A$. Similarly, user 2 can also correctly decode all the $2N$ bits of the desired message $B$.

%-----------------------------------------------------
For $(\theta_1,\theta_2)=(1,1)$, the following six coefficient matrices
\be
\label{Eq: full rank, (1,1), first point}
{\small
\begin{bmatrix}
\bm{\alpha}_{1,1}\\
\bm{\alpha}_{1,2}\\
\vdots\\
\bm{\alpha}_{1,N-1}\\
\gv_1
\end{bmatrix},
 \begin{bmatrix}
\bm{\alpha}_{2,1}\\
\bm{\alpha}_{2,2}\\
\vdots\\
\bm{\alpha}_{2,N-1}\\
\gv_3
\end{bmatrix},
 \begin{bmatrix}
\bm{\beta}_{1,1}\\
\bm{\beta}_{1,2}\\
\vdots\\
\bm{\beta}_{1,N-1}\\
\gv_2
\end{bmatrix},
 \begin{bmatrix}
\bm{\beta}_{2,1}\\
\bm{\beta}_{2,2}\\
\vdots\\
\bm{\beta}_{2,N-1}\\
\gv_4
\end{bmatrix},
\begin{bmatrix}
\uv_{1,1}\\
\uv_{2,1}\\
\vdots\\
\uv_{N-1,1}\\
\gv_1
\end{bmatrix},
\begin{bmatrix}
\uv_{1,2}\\
\uv_{2,2}\\
\vdots\\
\uv_{N-1,2}\\
\gv_3
\end{bmatrix},
}
\ee
are required to be full-rank. The alignment condition is
\begin{subequations}
\label{Eq: alignment, (1,1), first point}
\begin{align}
&\gv_2=\vv_{1,1}=\vv_{2,1}=\cdots=\vv_{N-1,1},\label{Eq: alignment g2, (1,1), first point}  \\
&\gv_4=\vv_{1,2}=\vv_{2,2}=\cdots=\vv_{N-1,2}.\label{Eq: alignment g4, (1,1), first point}
\end{align}
\end{subequations}
With the above full-rank and alignment conditions, we show that both users can correctly decode message $A$.

Due to the alignment of
(\ref{Eq: alignment g2, (1,1), first point}), we have $A_{N,2}^{[(1,1)]}=\gv_2B_{(1:N)}^{\rm T}=\vv_{1,1}B_{(1:N)}^{\rm T}=\cdots=\vv_{N-1,1}B_{(1:N)}^{\rm T}$. Subtracting $A_{N,2}^{[(1,1)]}$ from $A_{n,1}^{[(1,1)]},\forall n\in[N-1]$, and due to the full-rank condition of (\ref{Eq: full rank, (1,1), first point}), the users can solve $A_{(1:N)}$ from
\be
A_{(1:N)}^{\rm T}=
\begin{bmatrix}
\uv_{1,1}\\
\uv_{2,1}\\
\vdots\\
\uv_{N-1,1}\\
\gv_1
\end{bmatrix}^{-1}
\begin{bmatrix}
A_{1,1}^{[(1,1)]}-A_{N,2}^{[(1,1)]}\\
A_{2,1}^{[(1,1)]}-A_{N,2}^{[(1,1)]}\\
\vdots\\
A_{N-1,1}^{[(1,1)]}-A_{N,2}^{[(1,1)]}\\
A_{N,1}^{[(1,1)]}
\end{bmatrix}.
\ee
Also, due to the alignment of (\ref{Eq: alignment g4, (1,1), first point}), we have $A_{N,4}^{[(1,1)]}=\gv_4B_{(N+1:2N)}^{\rm T}=\vv_{1,2}B_{(N+1:2N)}^{\rm T}=\cdots=\vv_{N-1,2}B_{(N+1:2N)}^{\rm T}$. Subtracting $A_{N,4}^{[(1,1)]}$ from $A_{n,2}^{[(1,1)]},\forall n\in[N-1]$, and due to the full-rank condition of (\ref{Eq: full rank, (1,1), first point}), the users can solve $A_{(N+1:2N)}$ from
\be
A_{(N+1:2N)}^{\rm T}=
\begin{bmatrix}
\uv_{1,2}\\
\vdots\\
\uv_{N-1,2}\\
\gv_3
\end{bmatrix}^{-1}
\begin{bmatrix}
A_{1,2}^{[(1,1)]}-A_{N,4}^{[(1,1)]}\\
\vdots\\
A_{N-1,2}^{[(1,1)]}-A_{N,4}^{[(1,1)]}\\
A_{N,3}^{[(1,1)]}
\end{bmatrix}.
\ee
Therefore, both users can correctly decode message $A$.
\begin{remark}
\emph{Note that for the case for identical demands, i.e., $(\theta_1,\theta_2)=(1,1),(2,2)$, the cached contents of the users are actually not used in the decoding process.}
\end{remark}

With the above full-rank and alignment conditions, we now employ a randomized specification of the linear combination coefficients used by each DB and formally describe the delivery scheme.

We first introduce some necessary notations. Let $\mathbf{Y}_N \in\mathbb{F}_2^{N\times N}(N\ge 2)$ be a full-rank binary matrix defined as
$\mathbf{Y}_N\eqdef
\begin{bmatrix}
\mathbf{I}_{N-1} & \mathbf{1}_{N-1}^{\rm T}\\
\mathbf{0}_{N-1} & 1
\end{bmatrix}$.
Let $\mathcal{Y}_N$ be a set containing the rows of $\mathbf{Y}_N$, i.e., $\mathcal{Y}_N$ contains $N$ linearly independent binary row vectors. Also define two binary matrices $\mathbf{M}(\uv_{:,i},\gv_j),\mathbf{M}(\vv_{:,i},\gv_j)\in\mathbb{F}_2^{N\times N},\forall i\in[2],\forall j\in[4]$
as
$\mathbf{M}(\uv_{:,i},\gv_j)\triangleq \left[\uv_{1,i};\uv_{2,i};\cdots;\uv_{N-1,i};\gv_j \right]$,
 $\mathbf{M}(\vv_{:,i},\gv_j)\triangleq\left[\vv_{1,i};\vv_{2,i};\cdots;\vv_{N-1,i};\gv_j \right]$.
The private delivery schemes for different 
$(\theta_1,\theta_2)$ are given by
\begin{itemize}
\item[1)]{$\underline{(\theta_1,\theta_2)=(1,2)}$:} Let $\gv_1$ and $\gv_4$ be chosen randomly and uniformly i.i.d. from the rows in $\mathcal{Y}_N$. Then let $\mathbf{M}(\uv_{:,2},\gv_3)$ and $\mathbf{M}(\vv_{:,1},\gv_2)$ be two independent random permutations of the rows of $\mathbf{Y}_N$. It can be seen that with such a specification of the answer linear coefficients and the previously defined cache coefficients, the full-rank condition of (\ref{Eq: full rank (1,2), first point}) can be satisfied. Therefore, both users can correctly decode their desired messages.
\item[2)]{$\underline{(\theta_1,\theta_2)=(1,1)}$:} Let $\gv_2$ and $\gv_4$ be chosen randomly and uniformly i.i.d. from the rows in $\mathcal{Y}_N$. Then let $\mathbf{M}(\uv_{:,1},\gv_1)$ and $\mathbf{M}(\uv_{:,2},\gv_3)$ be two independent random permutations of the rows of $\mathbf{Y}_N$. It can be seen that the full-rank condition (\ref{Eq: full rank, (1,1), first point}) is satisfied.
\item[3)]{$\underline{(\theta_1,\theta_2)=(2,1)}$:} Let $\gv_2$ and $\gv_3$ be chosen randomly and uniformly i.i.d. from the rows in $\mathcal{Y}_N$. Then let $\mathbf{M}(\uv_{:,1},\gv_1)$ and $\mathbf{M}(\uv_{:,2},\gv_4)$ be two independent random permutations of the rows of $\mathbf{Y}_N$. It can be easily checked that the corresponding full-rank condition is satisfied.
\item[4)]{$\underline{(\theta_1,\theta_2)=(2,2)}$:} Let $\gv_1$ and $\gv_3$ be chosen randomly and uniformly i.i.d. from the rows in $\mathcal{Y}_N$. Then let $\mathbf{M}(\vv_{:,1},\gv_2)$ and $\mathbf{M}(\vv_{:,2},\gv_4)$ be two independent random permutations of the rows of $\mathbf{Y}_N$. It can be easily checked that the corresponding full-rank condition is satisfied.
\end{itemize}

We next prove the correctness and privacy of the above delivery scheme.

\emph{Correctness:} For any demands $(\theta_1,\theta_2)$, the random specification of the answer linear coefficients satisfies the corresponding full-rank and alignment conditions, implying the decodability.

\emph{Privacy:} WLOG, we prove that the above delivery scheme is private from DB 1's perspective, i.e., the demand vector $\bm{\theta}$ is equally likely to be $(1,2),(2,1),(1,1)$ or $(2,2)$. More specifically,  let $x\triangleq[\uv_{1,1},\uv_{1,2},\vv_{1,1},\vv_{1,2}]\in \mathcal{Y}_N^4$ be a random realization of the answer linear coefficients of DB 1. Let ${\Gamma}(\uv_{1,j},\bm{\theta})$ denote a random query of the value of $\uv_{1,j},j=1,2$ to DB 1 when the user demand vector is $\bm{\theta}$. Other notations follow similarly. Let
$
X(\bm{\theta})\triangleq[\Gamma(\uv_{1,1},\bm{\theta}),\Gamma(\uv_{1,2},\bm{\theta}),\Gamma(\vv_{1,1},\bm{\theta}),\Gamma(\vv_{1,2},\bm{\theta})]
$
represent the random query to DB 1 when the user demand vector is $\bm{\theta}$. Then the probability that $x$ is generated for $\bm{\theta}$ (i.e., $X(\bm{\theta})=x$) is
\begin{align}
\label{eq: privacy 1}
P(X(\bm{\theta})=x) &= P(\Gamma(\uv_{1,1},\bm{\theta})=\uv_{1,1})\times P(\Gamma(\uv_{1,2},\bm{\theta})=\uv_{1,2})\nonumber\\
&\times P(\Gamma(\vv_{1,1},\bm{\theta})=\vv_{1,1})\times P(\Gamma(\vv_{1,2},\bm{\theta})=\vv_{1,2}) \nonumber\\
&=\begin{cases}
& \frac{1}{N}\times \frac{(N-1)!}{N!}\times \frac{(N-1)!}{N!}\times \frac{1}{N}= \frac{1}{N^4},\quad \textrm{if $\bm{\theta}=(1,2)$}\\
&\frac{(N-1)!}{N!}\times\frac{1}{N}\times\frac{1}{N}\times \frac{(N-1)!}{N!}= \frac{1}{N^4},\quad \textrm{if $\bm{\theta}=(2,1)$}\\
& \frac{(N-1)!}{N!}\times\frac{(N-1)!}{N!}\times \frac{1}{N}\times\frac{1}{N}= \frac{1}{N^4},\quad \textrm{if $\bm{\theta}=(1,1)$}\\
& \frac{1}{N}\times\frac{1}{N}\times \frac{(N-1)!}{N!}\times\frac{(N-1)!}{N!}= \frac{1}{N^4},\quad \textrm{if $\bm{\theta}=(2,2)$}
\end{cases}. %\\
%&= \frac{1}{N^4}
\end{align}
Since $P(X(\bm{\theta})=x)$ does not depend on $\bm{\theta}$, from DB 1's perspective, the coefficient realization $x$ is equally likely to be generated for $\bm{\theta}=(1,2),(2,1),(1,1)$ or $(2,2)$. Therefore, the scheme is private from DB 1's point of view. Due to symmetry, the scheme is also private from any other individual DB's perspective. As a result, the proposed delivery scheme is private.

\emph{Performance:} Since $D=2N+2$ linear combinations, each containing one bit, are downloaded in total, the achieved load is $R=\frac{D}{L}=\frac{N+1}{N}$.

{We provide the following example to briefly illustrate how we choose the coefficients in the above proposed scheme.
 \begin{example}\label{example: N=2, (1/4,3/2)} (Achievability of $(\frac{1}{4},\frac{3}{2})$ for $N=2$)
Consider the cache-aided MuPIR problem with $K=2$ messages, $K_{\rm u}=2$ users and $N=2$ DBs.

\emph{1) Cache placement:} Assume that each message consists of $L=4$ bits, i.e., $A=(A_1,A_2,A_3,A_4)$, $B=(B_1,B_2,B_3,B_4)$. Each user stores a linear combination of the message bits which are
$Z_1 = \bm{\alpha}_{1,1} [A_1,A_2]^{\rm T} + \bm{\beta}_{1,1} [B_1,B_2]^{\rm T}  =A_1 +B_1$\footnote{{ With a slight abuse of notation, here we use $Z_1,Z_2$ to denote the cached bits of the users despite they are defined as sets.}} and
$Z_2 = \bm{\alpha}_{2,1} [A_3,A_4]^{\rm T} + \bm{\beta}_{2,1} [B_3,B_4]^{\rm T}  =A_3 +B_3$, i.e., we let $\bm{\alpha}_{1,1}=\bm{\alpha}_{2,1}=\bm{\beta}_{1,1}=\bm{\beta}_{2,1}=[1,0]$.
It can be seen that the cache memory constraint $M=\frac{1}{4}$ is satisfied.

\emph{2) Private delivery:}
For any demand vector, we construct the answers of the DBs as
\begin{equation}\label{Eq: answer matrix (1/4,3/2)}
{\small
\begin{bmatrix}
A_{1,1}^{[\bm{\theta}]}\\
A_{1,2}^{[\bm{\theta}]}\\
A_{2,1}^{[\bm{\theta}]}\\
A_{2,2}^{[\bm{\theta}]}\\
A_{2,3}^{[\bm{\theta}]}\\
A_{2,4}^{[\bm{\theta}]}
\end{bmatrix}=\begin{bmatrix}
\uv_{1,1}  & \bm{0}_2 & \vv_{1,1}  & \bm{0}_2\\
 \bm{0}_2 & \uv_{1,2} & \bm{0}_2  & \vv_{1,2}\\
\gv_1  & \bm{0}_2 &  \bm{0}_2  & \bm{0}_2\\
\bm{0}_2  & \bm{0}_2 & \gv_2  & \bm{0}_2\\
\bm{0}_2  & \gv_3 &\bm{0}_2 & \bm{0}_2\\
\bm{0}_2  & \bm{0}_2 & \bm{0}_2  &\gv_4\\
\end{bmatrix}\begin{bmatrix}
A_{(1:2)}^{\rm T}\\
A_{(3:4)}^{\rm T}\\
B_{(1:2)}^{\rm T}\\
B_{(3:4)}^{\rm T}
\end{bmatrix}.}
\end{equation}

Now suppose the demand vector is $(\theta_1,\theta_2)=(1,2)$.
For this demand vector, we let $\uv_{1,1}= \gv_1$ and $\vv_{1,2}= \gv_4$ as  shown in~\eqref{Eq: alignment, (1,2), first point}. In order to choose the coefficients in~\eqref{Eq: answer matrix (1/4,3/2)}, we introduce the matrix
$\mathbf{Y}_2= [1,1; 0,1]$ which is independent of the demand vector. Since the demand vector is  $(\theta_1,\theta_2)=(1,2)$,  we let $\gv_1$ and $\gv_4$ be chosen randomly and uniformly i.i.d. from $\mathcal{Y}_2=\{[1,1],[0,1]\}$.
In addition, we let $\left[\uv_{1,2}; \gv_3 \right]$ and $\left[\vv_{1,1}; \gv_2 \right]$ be two  independent random permutations of the rows of $\mathbf{Y}_2$. The purpose of the above coefficient selection is such that the message bits $A_3, A_4$ and $B_1, B_2$ can be directly decoded  from the answers by both users without using the cache. For user 1, the interference of $B_1, B_2$ will be eliminated and a linear combination of $A_1$ and $A_2$ is left. Together with the one aligned equation of $A_1$ and $A_2$  from the answers, user 1 can decode $A_1$ and $A_2$. Similarly, user 2 can eliminate the interference of $A_3$ and $A_4$, left with a linear combination of $B_3$ and $B_4$ in the cache. Together with the one received equation of $B_3$ and $B_4$, user can decode $B_3$ and $B_4$.

\emph{Correctness:}
From $A_{1,2}^{[\bm{\theta}]}- A_{2,4}^{[\bm{\theta}]}$, each user can decode
$\uv_{1,2} [A_3,A_4]^{\rm T}$.
In addition, from $A_{2,3}^{[\bm{\theta}]}$, each user receives $\gv_3 [A_3,A_4]^{\rm T}$. Since  $ \uv_{1,2}$ and  $\gv_3 $  are two different rows of $\mathbf{Y}_2$, we can see they are linearly independent. Thus each user can decode $A_3$ and $A_4$.
From $A_{1,1}^{[\bm{\theta}]}- A_{2,1}^{[\bm{\theta}]}$, each user can decode   $
\vv_{1,1} [B_1,B_2]^{\rm T}
$. In addition, from $A_{2,2}^{[\bm{\theta}]}$, each user receives $\gv_2 [B_1,B_2]^{\rm T}$. Since  $ \vv_{1,1}$ and  $\gv_2 $  are two different rows of $\mathbf{Y}_2$, we can see they are linearly independent. Thus each user can decode $B_1$ and $B_2$. Since user $1$ caches $A_1 +B_1$  and has decoded $B_1$, it can then recover $A_1$. From $A_{2,1}^{[\bm{\theta}]}$, user $1$ receives $\gv_1 [A_1,A_2]^{\rm T}$ where $\gv_1$ is one row in $\mathcal{Y}_2$. No matter which row is $\gv_1$, user $1$ can always recover $A_1$ and $A_2$ from $\gv_1 [A_1,A_2]^{\rm T}$. Hence, user $1$ can recover the whole file $A$. Similarly, since user $2$ caches $A_3 +B_3$ and has decoded $A_3$, it can then recover $B_3$. From $A_{2,4}^{[\bm{\theta}]}$, user $2$ receives $\gv_4 [B_3,B_4]^{\rm T}$ where $\gv_4$ is one row in $\mathcal{Y}_2$. No matter which row is $\gv_4$, user $2$ can always recover $B_3$ and $B_4$ from $\gv_4 [B_3,B_4]^{\rm T}$. Hence, user $2$ can recover the whole file $B$.

\emph{Privacy:}
Intuitively,  from the viewpoint of DB 1  whose sent linear combinations are $\gv_1 [A_1,A_2]^{\rm T} + \vv_{1,1} [B_1,B_2]^{\rm T}$ and $\uv_{1,2} [A_3,A_4]^{\rm T} + \gv_4 [B_3,B_4]^{\rm T}$, the vectors $\gv_1, \gv_4, \vv_{1,1}, \uv_{1,2}$ are  randomly and independently  chosen from the rows of $\mathbf{Y}_2$. Thus, the sent linear combinations are independent of the demand vector.
From the viewpoint of DB 2    whose sent linear combinations are $\gv_1 [A_1,A_2]^{\rm T}$, $\gv_2 [B_1,B_2]^{\rm T}$,
$\gv_3 [A_3,A_4]^{\rm T}$, and $\gv_4 [B_3,B_4]^{\rm T}$,
the vectors  $\gv_1, \gv_2, \gv_3, \gv_4$  are  randomly and independently  chosen from the rows of $\mathbf{Y}_2$. Thus, the sent linear combinations are independent of the demand vector.
Therefore, each  DB cannot get any information about the demand vector from its sent linear combinations and the user cache.

\emph{Performance:} The achieved load is $R=\frac{D}{L}=\frac{6}{4}=\frac{3}{2}$.
\hfill $\lozenge$
\end{example}
}

%----------------------------------------------------------------------------------
\subsection{Achievability of $\left(\frac{2(N-1)}{2N-1},\frac{N+1}{2N-1}\right)$}
%------------------- exmaple 2 --------------------
\if{0}
\begin{example}
\label{example: N=2, (2/3,1)}
(Achievability of $(\frac{2}{3},1)$ for $N=2$)
Consider the same setting as Example \ref{example: N=2, (1/4,3/2)} where $K=K_{\rm u}=N=2$. In this example we show the achievability of the pair $(\frac{2}{3},1)$ using
the idea of CIA.
The cache placement and private delivery phases are described as follows.

\emph{1) Cache placement:} Assume that each message consists of $L=3$ bits, i.e., $A=(A_1,A_2,A_3)$, $B=(B_1,B_2,B_3)$. Each user stores two message bits, i.e.,
$Z_1 = \{A_1,B_1 \}, Z_2 = \{A_2, B_2 \}$.

\emph{2) Private delivery:} We first present the construction of the answers and then describe the delivery scheme. The answers of the DBs consists of linear combinations (over binary field) of certain message bits. In particular, the answer of DB 1 consists of a random linear combination given by
$A_{1}^{[\bm{\theta}]}=\uv_1A_{(1:3)}^{\rm T}+\vv_1B_{(1:3)}^{\rm T}$,
in which the random linear combination coefficients $\uv_1\triangleq [u_{1,1},u_{1,2},u_{1,3}]$, $\vv_1\triangleq[v_{1,1},v_{1,2},v_{1,3}]\in\mathbb{F}_2^{1\times 3}\backslash \{[0,0,0]\}$ are subject to design according to the user demands $(\theta_1,\theta_2)$. The answer of DB 2 consists of two random linear combinations, $A_2^{[\bm{\theta}]}=(A_{2,1}^{[\bm{\theta}]},A_{2,2}^{[\bm{\theta}]})$ for which
$A_{2,1}^{[\bm{\theta}]} =\gv_1 A_{(1:3)}^{\rm T},
A_{2,2}^{[\bm{\theta}]} = \gv_2 B_{(1:3)}^{\rm T}$,
where the linear coefficients $\gv_1 \triangleq[g_{1,1},g_{1,2},g_{1,3}]$, $\gv_2 \triangleq[g_{2,1},g_{2,2},g_{2,3}]\in \mathbb{F}_2^{1\times 3} \backslash \{[0,0,0]\}$ are subject to design. The answers can be written in a more compact form as
\be
\label{eqn: answer in matrix form N=2 (2/3,1)}
\begin{bmatrix}
A_{1}^{[\bm{\theta}]}\\
A_{2,1}^{[\bm{\theta}]}\\
A_{2,2}^{[\bm{\theta}]}\\
\end{bmatrix} =\begin{bmatrix}
\uv_1 & \vv_1\\
\gv_1  & \bm{0}_3\\
\bm{0}_3  & \gv_2
\end{bmatrix}\begin{bmatrix}
A_{(1:3)}^{\rm T}\\
B_{(1:3)}^{\rm T}
\end{bmatrix}.
\ee
In the delivery phase, the users download the two answers from the DBs. To guarantee correct decoding and privacy, the full-rank conditions and alignment conditions should be satisfied. We next show that how the linear coefficients can be designed
given the user demands. Similar to before, we will illustrate the case of $(\theta_1,\theta_2)\ = (1,2)$.

Let $(\theta_1,\theta_2)=(1,2)$. The following two coefficient matrices are required to be full-rank
\be \label{Eq: full rank [1,2], (2/3,1)}
\textrm{\emph{Full-rank condition}:}\quad
\begin{bmatrix}
u_{1,2} & u_{1,3}\\
g_{1,2} & g_{1,3}
\end{bmatrix},
\begin{bmatrix}
v_{1,1} & v_{1,3}\\
g_{2,1} & g_{2,3}
\end{bmatrix},
\ee
and the alignment is
\begin{eqnarray}
\label{Eq: alignment [1,2], (2/3,1)}
\textrm{\emph{Alignment condition}:}\quad
\,[g_{1,1},g_{1,3}]= [u_{1,1},u_{1,3}],\;
[g_{2,2},g_{2,3}] = [v_{1,2},v_{1,3}].
\end{eqnarray}
The alignment condition implies that the message bits $A_1, A_3$ are aligned among the equations $A_1^{[(1,2)]}$ and $A_{2,1}^{[(1,2)]}$, and $B_2, B_3$ are aligned among $A_1^{[(1,2)]}$ and $A_{2,2}^{[(1,2)]}$. We highlight the purpose of such an alignment as follows. For user 1, since $A_1$ is already in its cache, it needs to recover $A_2$ and $A_3$ from the received %equations (i.e., linear combinations).
linear combinations. The alignment of $B_2$ and $B_3$ among $A_1^{[(1,2)]}$ and $A_{2,2}^{[(1,2)]}$ allows user 1 to remove the interference of $B_2$ and $B_3$ from equation $A_{1}^{[(1,2)]}$ (note that $B_1$ is cached). Due to the full-rank condition $\textrm{rank}([u_{1,2},u_{1,3};v_{1,2},v_{1,3}])=2$, user 1 now obtains two linearly independent equations $A_{1}^{[(1,2)]},A_{2,1}^{[(1,2)]}$ with unknowns $A_2$ and $A_3$, from which the two desired bits $A_2,A_3$ can be recovered. This decoding process can be summarized as
\be
\begin{bmatrix}
A_2\\
A_3
\end{bmatrix} =\begin{bmatrix}
u_{1,2} & u_{1,3}\\
g_{1,2} & g_{1,3}
\end{bmatrix}^{-1}
\left(\begin{bmatrix}
 A_1^{[(1,2)]}\\
  A_{2,1}^{[(1,2)]}
 \end{bmatrix}-\begin{bmatrix}
  A_{2,2}^{[(1,2)]} -g_{2,1}B_1\\
  0
 \end{bmatrix}-\begin{bmatrix}
 u_{1,1} \\
 g_{1,1}
 \end{bmatrix}A_1-\begin{bmatrix}
 v_{1,1} \\
 0
 \end{bmatrix}B_1\right).
\ee
Similarly, for user 2, since $B_2$ is cached, it needs to recover $B_1,B_3$ from the received equations. The alignment of $A_1,A_3$ among $A_1^{[(1,2)]}$ and $A_{2,2}^{[(1,2)]}$ allows user 2 to remove the interference of $A_1$ and $A_3$ from $A_{1}^{[(1,2)]}$ ($A_2$ is cached). The full-rank condition $\textrm{rank}([v_{1,1},v_{1,3};g_{2,1},g_{2,3}])=2$ ensures that user 2 obtains two linearly independent equations $A_1^{[(1,2)]},A_{2,2}^{[(1,2)]}$ with two unknowns $B_1$ and $B_3$, from which the two desired bits $B_1,B_3$ can be recovered, i.e.,
\be
\begin{bmatrix}
B_1\\
B_3
\end{bmatrix} =\begin{bmatrix}
v_{1,1} & v_{1,3}\\
g_{2,1} & g_{2,3}
\end{bmatrix}^{-1}
\left(\begin{bmatrix}
 A_1^{[(1,2)]}\\
  A_{2,2}^{[(1,2)]}
 \end{bmatrix}-\begin{bmatrix}
  A_{2,1}^{[(1,2)]} -g_{2,1}A_2\\
  0
 \end{bmatrix}-\begin{bmatrix}
 u_{1,2} \\
0
 \end{bmatrix}A_2-\begin{bmatrix}
 u_{1,2} \\
 g_{1,2}
 \end{bmatrix}B_2  \right).
\ee

With the above
constraints on the linear coefficients, we have the following observation.

\begin{observation}\emph{
Due to the full-rank and alignment conditions corresponding to each $(\theta_1,\theta_2)\in[2]^2$, it must hold that $u_{1,3}=v_{1,3}=1$ and $g_{1,3}=g_{2,3}=1$. More specifically, for $(\theta_1,\theta_2)=(1,2)$, from (\ref{Eq: alignment [1,2], (2/3,1)}), we have $u_{1,3}=g_{1,3},v_{1,3}=g_{2,3}$. In order for the two matrices in (\ref{Eq: full rank [1,2], (2/3,1)}) to be full-rank, then we must have $u_{1,3}=g_{1,3}=1$, $v_{1,3}=g_{2,3}=1$. For the sake of privacy, this condition should be satisfied for all possible user demands. Therefore, there are $2^4=16$ choices for choosing the DB 1 coefficients $[u_{1,1},u_{1,2},v_{1,1},v_{1,2}]\in\mathbb{F}_2^{1\times 4}$ and also $2^4$ choices for choosing the DB 2 coefficients $[g_{1,1},g_{1,2},g_{2,1},g_{2,2}]\in\mathbb{F}_2^{1\times 4}$. For each choice of DB 1 coefficients $[u_{1,1},u_{1,2},v_{1,1},v_{1,2}]$, the DB 2 coefficients $[g_{1,1},g_{1,2},g_{2,1},g_{2,2}]$ can be uniquely determined for each $(\theta_1,\theta_2)$ according to the full-rank and alignment conditions therein. Reversely, for each choice of the DB 2 coefficients $[g_{1,1},g_{1,2},g_{2,1},g_{2,2}]$, the DB 1 coefficients $[u_{1,1},u_{1,2},v_{1,1},v_{1,2}]$ can also be uniquely determined for each $(\theta_1,\theta_2)$.}
\end{observation}

With this observation, we are now ready to present the delivery scheme via a randomized specification of the linear coefficients. We first define two matrices as follows:
\be
\label{Eq: candidate matrix, (2/3,1)}
\mathbf{D}_1 \triangleq
\begin{bmatrix}
 0 & 0 &1 & 0 & 0 &1\\
  1 & 1 &1 & 1 & 1 &1\\
  1 & 0 &1 & 1 & 0 &1\\
    0 & 1 &1 & 0 & 1 &1
\end{bmatrix},\;
\mathbf{D}_2 \triangleq
\begin{bmatrix}
 0 & 1 &1 & 1 & 0 &1\\
  1 & 0 &1 & 0 & 1 &1\\
   1 & 1 &1 & 0 & 0 &1\\
    0 & 0 &1 & 1 & 1 &1
\end{bmatrix}.
\ee
Each row of $\mathbf{D}_1$ represents a realization of the DB 1 answer linear coefficients $[\uv_1,\vv_1]$ and each row of $\mathbf{D}_2$ represents a realization of the DB 2 answer coefficients $[\gv_1,\gv_2]$. With this construction, the delivery phase works as follows.

The users download an answer from each DB, and the linear coefficients are determined in the following way:
Let the DB 1 coefficient vector $[\uv_1,\vv_1]$ be chosen randomly from the rows of $\mathbf{D}_1$ with equal probabilities. Depending on the DB 1 choice and user demands, the DB 2 coefficient vector $[\gv_1,\gv_2]$ can be determined. For example,
when $(\theta_1,\theta_2)=(1,2)$,  if $\mathbf{D}_1(1,:)$ is chosen, let $[\gv_1,\gv_2]=\mathbf{D}_2(1,:)$.

We next prove the correctness and privacy the above delivery scheme.

\emph{Correctness:} Decodabilility is straightforward since the construction of the linear coefficients in $\mathbf{D}_1$ and $\mathbf{D}_2$ follows the full-rank and alignment conditions for any user demands $(\theta_1,\theta_2)\in[2]^2$.

\emph{Privacy:} First, the above delivery phase is private from the perspective of DB 1. This is because for any of the DB 1 coefficient choice $[\uv_1,\vv_1]$, there exists four different configurations of the DB 2 coefficients $[\gv_1,\gv_2]$ each of which corresponds to a different user demand vector $(\theta_1,\theta_2)\in[2]^2$. Since DB does not know the choice of DB 2, it can not tell which $(\theta_1,\theta_2)$ is being requested by the users. Similarly, the delivery is also private from DB 2's perspective. As a result, the above delivery scheme is private.

\emph{Performance:} Since $D=3$ linear combinations each containing one bit are downloaded in total, the achieved load is $R=\frac{D}{L}=1$.
\hfill $\lozenge$
\end{example}
\fi

%---------------------------------------------------
\emph{1) Cache placement:} Let $W_1=A, W_2=B$ be the two messages each of which  consists of $L=2N-1$ bits, i.e., $A=(A_1,A_2,\cdots, A_{2N-1})$, $B=(B_1,B_2,\cdots, B_{2N-1})$.  Each user stores $2(N-1)$ uncoded bits from each message, i.e.,
\begin{eqnarray}
 Z_1 &=& \{A_{1:N-1},B_{1:N-1}\},\\
 Z_2 &=& \{A_{N:2N-2},B_{N:2N-2}\}.
\end{eqnarray} Note that neither of the two users store the message bits $A_{2N-1}$ and $B_{2N-1}$.

\emph{2) Private delivery:} We first construct the answers from the DBs. The answer of DB $n\in[N-1]$ is a linear combination of certain message bits,
$A_{n}^{[\bm{\theta}]}= \uv_{n}A_{(1:2N-1)}^{\rm T} + \vv_{n}B_{(1:2N-1)}^{\rm T}$.
The answer of DB $N$ consists of two linear combinations, i.e., $A_N^{[\bm{\theta}]}=(A_{N,1}^{[\bm{\theta}]},A_{N,2}^{[\bm{\theta}]})$  where
$A_{N,1}^{[\bm{\theta}]} = \gv_1 A_{(1:2N-1)}^{\rm T},
A_{N,2}^{[\bm{\theta}]} = \gv_2 B_{(1:2N-1)}^{\rm T}$.
The linear combination coefficients $\uv_n \triangleq[u_{n,1},u_{n,2},\cdots, u_{n,2N-1}]$, $\vv_n \triangleq[v_{n,1},v_{n,2},\cdots, v_{n,2N-1}],\forall n\in[N-1]$, $\gv_j \triangleq [g_{j,1},g_{j,2},\cdots, g_{j,2N-1}],j=1,2$,
belong to $\mathbb{F}_2^{1\times (2N-1)}\backslash\{\bm{0}_{2N-1}\} $ and are subject to design according to different user demands. These answers can be written in a more compact form as
{
\begin{equation}
\label{Eq: answer in matrix form, second point }
\begin{bmatrix}
A_1^{[\bm{\theta}]}\\
\vdots\\
A_{N-1}^{[\bm{\theta}]}\\
A_{N,1}^{[\bm{\theta}]}\\
A_{N,2}^{[\bm{\theta}]}
\end{bmatrix}= \begin{bmatrix}
\uv_1 & \vv_1\\
\vdots & \vdots \\
\uv_{N-1} & \vv_{N-1}\\
\gv_1  & \bm{0}_{2N-1}\\
\bm{0}_{2N-1} &  \gv_2
\end{bmatrix}\begin{bmatrix}
A_{(1:2N-1)}^{\rm T}\\
B_{(1:2N-1)}^{\rm T}
\end{bmatrix} =
\begin{bmatrix}
\uv_{1,(1:2N-2)} & 1 & \vv_{1,(1:2N-2)} &1 \\
\vdots & \vdots& \vdots& \vdots \\
\uv_{N-1,(1:2N-2)}  & 1 & \vv_{N-1,(1:2N-2)} & 1 \\
\gv_{1,(1:2N-2)}  & 1 & \bm{0}_{2N-2} & 0\\
\bm{0}_{2N-2} & 0 &  \gv_{2,(1:2N-2)} &1
\end{bmatrix}\begin{bmatrix}
A_{(1:2N-1)}^{\rm T}\\
B_{(1:2N-1)}^{\rm T}
\end{bmatrix},
\end{equation}
where we let
\begin{align}
\label{Eq: observation 3 constant, second point}
 g_{1,2N-1} = g_{2,2N-1}=
u_{n,2N-1}=v_{n,2N-1}=1,\quad \forall n\in[N-1].
\end{align}
}

We next consider the four possible demands $(\theta_1,\theta_2)$ and show that the full-rank and alignment conditions are necessary for correct decoding. For simplicity, we will focus on the cases of $(\theta_1,\theta_2)=(1,2)$ and $(\theta_1,\theta_2)=(1,1)$. The cases of $(\theta_1,\theta_2)=(2,1)$ and $(2,2)$ follow similarly.

%-------------------------------------------
For $(\theta_1,\theta_2)=(1,2)$, the following two coefficient matrices  are required to be full-rank, i.e.,
\be
\label{Eq: full rank, (1,2), second point}
\textrm{\emph{Full-rank condition}:}\quad
\begin{bmatrix}
\uv_{1,(N:2N-1)}\\
\vdots\\
\uv_{N-1,(N:2N-1)}\\
\gv_{1,(N:2N-1)}
\end{bmatrix},
\begin{bmatrix}
[\vv_{1,(1:N-1)},v_{1,2N-1}]\\
\vdots\\
[\vv_{N-1,(1:N-1)},v_{N-1,2N-1}]\\
[\gv_{2,(1:N-1)},g_{2,2N-1}]
\end{bmatrix}.
\ee
For alignment, we let $\forall n\in[N-1]$:
\begin{subequations}
\label{Eq: alignment, (1,2), second point}
\begin{align}
\textrm{\emph{Alignment condition}:}\quad
[\uv_{n,(1:N-1)},u_{n,2N-1}]&=[\gv_{1,(1:N-1)},g_{1,2N-1}],
\label{Eq: alignment g1, (1,2), second point}\\
\vv_{n,(N:2N-1)}&= \gv_{2,(N:2N-1)},
\label{Eq: alignment g2, (1,2), second point}
\end{align}
\end{subequations}
i.e., the message bits $A_{(1:N-1)},A_{2N-1}$ are aligned among the linear combinations $A_1^{[(1,2)]}$, $A_2^{[(1,2)]}$, $\cdots$, $A_{N-1}^{[(1,2)]}$, and $A_{N,1}^{[(1,2)]}$ and the bits $B_{(N:2N-1)}$ are aligned among the linear combinations $A_1^{[(1,2)]}$, $A_2^{[(1,2)]},\cdots, A_{N-1}^{[(1,2)]}$, and $A_{N,2}^{[(1,2)]}$. We next show that the above two conditions guarantee correct decoding.

For user 1, due to the alignment of $B_{(N:2N-1)}$, we have
\be
A_{N,2}^{[(1,2)]}- \gv_{2,(1:N-1)}B_{(1:N-1)}^{\rm T}=\gv_{2,(N:2N-1)}B_{(N:2N-1)}^{\rm T}=\vv_{n,(N:2N-1)}  B_{(N:2N-1)}^{\rm T},  \forall n\in[N-1].
\ee
Subtracting $A_{N,2}^{[(1,2)]}- \gv_{2,(1:N-1)}B_{(1:N-1)}^{\rm T}$ (this is known to user 1 since $B_{1:N-1}$ are already cached by user 1) from $A_1^{[(1,2)]},A_2^{[(1,2)]},\cdots,A_{N-1}^{[(1,2)]}$ in {(\ref{Eq: answer in matrix form, second point })}, together with $A_{N,1}^{[(1,2)]}=\gv_1A_{(1:2N-1)}^{\rm T}$,  we obtain $N$ independent linear combinations of $A_{(N:2N-1)}$,
which can be solved as
\be
A_{(N:2N-1)}^{\rm T}=
\begin{bmatrix}
\uv_{1,(N:2N-1)}\\
\vdots\\
\uv_{N-1,(N:2N-1)}\\
\gv_{1,(N:2N-1)}
\end{bmatrix}^{-1}\yv,
\ee
in which
\be
{\small
\yv \triangleq\begin{bmatrix}
A_1^{[(1,2)]}-(A_{N,2}^{[(1,2)]}-\gv_{2,(1:N-1)}B_{(1:N-1)}^{\rm T})\\
\vdots\\
A_{N-1}^{[(1,2)]}-(A_{N,2}^{[(1,2)]}-\gv_{2,(1:N-1)}B_{(1:N-1)}^{\rm T})\\
A_{N,1}^{[(1,2)]}
\end{bmatrix}
-\begin{bmatrix}
\uv_{1,(1:N-1)} &  \vv_{1,(1:N-1)}\\
\vdots\\
\uv_{N-1,(1:N-1)} &  \vv_{N-1,(1:N-1)}\\
\gv_{1,(1:N-1)} & \bm{0}_{N-1}
\end{bmatrix}
\begin{bmatrix}
A_{(1:N-1)}^{\rm T}\\
B_{(1:N-1)}^{\rm T}
\end{bmatrix}.
}
\ee
Since the message bits $A_{(1:N-1)},B_{(1:N-1)}$ are already cached by user 1, it can decode the bits $A_{(N:2N-1)}$ and correctly  recover message $A$. Similarly, user 2 can decode the bits $B_{(1:N-1)},B_{2N-1}$ and correctly recover message $B$.

%---------------------------------------------
For $(\theta_1,\theta_2)=(1,1)$, the following two coefficient matrices
\be
\label{eqn: full rank, (1,1), second point}
{\small
\begin{bmatrix}
\uv_{1,(N:2N-1)}\\
\uv_{2,(N:2N-1)}\\
\vdots\\
\uv_{N-1,(N:2N-1)}\\
\gv_{1,(N:2N-1)}
\end{bmatrix},
\begin{bmatrix}
[\uv_{1,(1:N-1)},u_{1,2N-1}]\\
[\uv_{2,(1:N-1)},u_{2,2N-1}]\\
\vdots\\
[\uv_{N-1,(1:N-1)},u_{N-1,2N-1}]\\
[\gv_{1,(1:N-1)},g_{1,2N-1}]
\end{bmatrix},
}
\ee
are required to be full-rank and the alignment is:
\be
\label{Eq: alignment, (1,1), second point}
\gv_2  = \vv_n, \quad \forall n\in[N-1].
\ee
The decoding is explained as follows. Due to the alignment of (\ref{Eq: alignment, (1,1), second point}), we have
\be
A_{N,2}^{[(1,1)]}=\gv_2B_{(1:2N-1)}^{\rm T}=\vv_1B_{(1:2N-1)}^{\rm T} =\cdots=\vv_{N-1}B_{(1:2N-1)}^{\rm T}, \forall n\in[N-1].
\ee
Subtracting $A_{N,2}^{[(1,1)]}$ from all $A_{n}^{[(1,1)]},\forall n\in[N-1]$, we obtain
\be
{\small
A_{(N:2N-1)}^{\rm T}=
\begin{bmatrix}
\uv_{1,(N:2N-1)}\\
\uv_{2,(N:2N-1)}\\
\vdots\\
\uv_{N-1,(N:2N-1)}\\
\gv_{1,(N:2N-1)}
\end{bmatrix}^{-1}
\left(
\begin{bmatrix}
A_1^{[(1,1)]}-A_{N,2}^{[(1,1)]}\\
A_2^{[(1,1)]}-A_{N,2}^{[(1,1)]}\\
\vdots\\
A_{N-1}^{[(1,1)]}-A_{N,2}^{[(1,1)]}\\
A_{N,1}^{[(1,1)]}
\end{bmatrix}-
\begin{bmatrix}
\uv_{1,(1:N-1)}\\
\uv_{2,(1:N-1)}\\
\vdots\\
\uv_{N-1,(1:N-1)}\\
\gv_{1,(1:N-1)}
\end{bmatrix}A_{(1:N-1)}^{\rm T}
\right).
}
\ee
Since the message bits $A_{1:N-1}$ are cached by user 1, it can decode the desired bits $A_{N:2N-1}$ and correctly recover $A$. Similarly, user 2 can decode the desired bits $A_{1:N-1},A_{2N-1}$. Therefore, both users can correctly recover the desired message $A$.

With the above full-rank and alignment conditions, we now employ a randomized specification of the linear combination coefficients used by each DB and formally describe the delivery scheme.

We first introduce some necessary notations.
Let $ \mathbf{Y}^{\prime}_{N}\triangleq  [\mathbf{I}_{N-1}; \mathbf{0}_{N-1}]$ be a binary matrix with dimension $N\times( N-1)$ and let  $\mathcal{Y}^{\prime}_N$ be a set containing the rows of $\mathbf{Y}^{\prime}_N$. For an index vector $(m:n)=(m,m+1,...,n-1,n)$ where $m\le n$, define $\forall j=1,2$:
\begin{subequations}
\begin{align}
&\mathbf{M}^{\prime}(\uv,\gv_j,(m:n))\triangleq\left[\uv_{1,(m:n)}; \uv_{2,(m:n)};\cdots; \uv_{N-1,(m:n)};\gv_{j,(m:n)}    \right]\in\mathbb{F}_2^{N\times (n-m+1)},\\
&\mathbf{M}^{\prime}(\vv,\gv_j,(m:n))\triangleq\left[\vv_{1,(m:n)}; \vv_{2,(m:n)};\cdots; \vv_{N-1,(m:n)};\gv_{j,(m:n)}    \right]\in\mathbb{F}_2^{N\times (n-m+1)}.
\end{align}
\end{subequations}
The delivery schemes for different $(\theta_1,\theta_2)$ are then as follows.
\begin{itemize}
\item[1)]{\underline{$(\theta_1,\theta_2)=(1,2)$}:} Let $\gv_{1,(1:N-1)}$ and $\gv_{2,(N:2N-2)}$ be chosen randomly and uniformly i.i.d. from the rows in $\mathcal{Y}^{\prime}_N$. Also, let $\mathbf{M}^{\prime}(\uv,\gv_1,(N:2N-2))$ and $\mathbf{M}^{\prime}(\vv,\gv_2,(1:N-1))$ be two independent random permutations of the rows of $\mathbf{Y}^{\prime}_N$. It can be easily seen that the full-rank condition of (\ref{Eq: full rank, (1,2), second point}) is satisfied, implying the decodability.
\item[2)]{\underline{$(\theta_1,\theta_2)=(1,1)$}:} Let $\gv_{2,(N:2N-2)}$ and $\gv_{2,(1:N-1)}$ be chosen randomly and uniformly i.i.d. from the rows in $\mathcal{Y}^{\prime}_N$. Also, let $\mathbf{M}^{\prime}(\uv,\gv_1,(N:2N-2))$ and $\mathbf{M}^{\prime}(\uv,\gv_1,(1:N-1))$ be two independent random permutations of the rows of $\mathbf{Y}^{\prime}_N$.
\item[3)]{\underline{$(\theta_1,\theta_2)=(2,1)$}:} Let $\gv_{1,(N:2N-2)}$ and $\gv_{2,(1:N-1)}$ be chosen randomly and uniformly i.i.d. from the rows in $\mathcal{Y}^{\prime}_N$. Also, let $\mathbf{M}^{\prime}(\uv,\gv_1,(1:N-1))$ and $\mathbf{M}^{\prime}(\vv,\gv_2,(N:2N-2))$ be two independent random permutations of the rows of $\mathbf{Y}^{\prime}_N$.
\item[4)]{\underline{$(\theta_1,\theta_2)=(2,2)$}:} Let $\gv_{1,(N:2N-2)}$ and $\gv_{1,(1:N-1)}$ be chosen randomly and uniformly i.i.d. from the rows in $\mathcal{Y}^{\prime}_N$. Also, let $\mathbf{M}^{\prime}(\vv,\gv_2,(1:N-1))$ and $\mathbf{M}^{\prime}(\vv,\gv_2,(N:2N-2))$ be two independent random permutations of the rows of $\mathbf{Y}^{\prime}_N$.
\end{itemize}

\emph{Correctness:} Decodability is straightforward since the randomized specifications of the linear coefficients for different user demands guarantee the corresponding full-rank and alignment conditions.

\emph{Privacy:} By the similar argument in (\ref{eq: privacy 1}), it can be seen that this scheme is private.

\emph{Performance:} Since $D=N+1$ linear combinations, each containing one bit, are downloaded in total, the achieved load is $R=\frac{D}{L}=\frac{N+1}{2N-1}$.

{
We provide the following example to briefly illustrate how we choose the coefficients in the above proposed scheme.
\begin{example}
\label{example: N=2, (2/3,1)}
(Achievability of $(\frac{2}{3},1)$ for $N=2$)
Consider the same setting as Example \ref{example: N=2, (1/4,3/2)} where $K=K_{\rm u}=N=2$. In this example we show the achievability of the memory-load pair $(\frac{2}{3},1)$.

\emph{1) Cache placement:} Assume that each message consists of $L=3$ bits, i.e., $A=(A_1,A_2,A_3)$, $B=(B_1,B_2,B_3)$. Each user stores two message bits, i.e.,
$Z_1 = \{A_1,B_1 \}, Z_2 = \{A_2, B_2 \}$.

\emph{2) Private delivery:} For any demand vector, the answers of the DBs are constructed as
\begin{equation}
\label{Eq: ex answer in matrix form, second point}
\begin{bmatrix}
A_1^{[\bm{\theta}]}\\
A_{2,1}^{[\bm{\theta}]}\\
A_{2,2}^{[\bm{\theta}]}
\end{bmatrix}= \begin{bmatrix}
\uv_1 & \vv_1\\
\gv_1  & \bm{0}_{3}\\
\bm{0}_{3} &  \gv_2
\end{bmatrix}\begin{bmatrix}
A_{(1:3)}^{\rm T}\\
B_{(1:3)}^{\rm T}
\end{bmatrix}= \begin{bmatrix}
u_{1,1} & u_{1,2} & 1 & v_{1,1} & v_{1,2} & 1 \\
g_{1,1} &g_{1,2} &1 & 0 & 0 & 0\\
 0 & 0 & 0 & g_{2,1} &g_{2,2} &1
\end{bmatrix}\begin{bmatrix}
A_{(1:3)}^{\rm T}\\
B_{(1:3)}^{\rm T}
\end{bmatrix}.
\end{equation}

Now consider $(\theta_1,\theta_2)=(1,2)$.
For this demand vector, we let $ u_{1,1}=g_{1,1}$, $v_{1,2}= g_{2,2}$, as shown in~\eqref{Eq: alignment, (1,2), second point}. Thus the answers of the DBs in~\eqref{Eq: ex answer in matrix form, second point} become
\begin{equation}
\label{Eq: ex answer in matrix form, second point 2}
\begin{bmatrix}
A_1^{[\bm{\theta}]}\\
A_{2,1}^{[\bm{\theta}]}\\
A_{2,2}^{[\bm{\theta}]}
\end{bmatrix}  = \begin{bmatrix}
g_{1,1} & u_{1,2} & 1 & v_{1,1} & g_{2,2} & 1 \\
g_{1,1} &g_{1,2} &1 & 0 & 0 & 0\\
 0 & 0 & 0 & g_{2,1} &g_{2,2} &1
\end{bmatrix}\begin{bmatrix}
A_{(1:3)}^{\rm T}\\
B_{(1:3)}^{\rm T}
\end{bmatrix}.
\end{equation}

 In order to specify the coefficients in~\eqref{Eq: ex answer in matrix form, second point},
 we choose the matrix
$\mathbf{Y}^{\prime}_2\triangleq
[1,0]^{\rm T}$ and thus $\mathcal{Y}_2^{\prime}=\{0,1\}$, both of which are independent of the demand vector. Since the    demand vector is $(\theta_1,\theta_2)=(1,2)$, we let $g_{1,1}$ and $g_{2,2}$ be chosen randomly and uniformly i.i.d. from  $\mathcal{Y}^{\prime}_2$.
 Also, we let $[u_{1,2};g_{1,2} ]$ and  $[v_{1,1};g_{2,1}]$  be two independent  random permutations of the rows of $\mathbf{Y}^{\prime}_2$.

 \emph{Correctness:}
 We first focus on user $1$ who caches $A_1, B_1$ and desires message $A$.
 From $ A_{2,1}^{[\bm{\theta}]}$ and its cached content $A_1$, user $1$   decodes $g_{1,2} A_2  + A_3$.
 From $A_1^{[\bm{\theta}]}-A_{2,2}^{[\bm{\theta}]}$, user $1$   decodes
$g_{1,1} A_1 +u_{1,2} A_2 +   A_3 +(v_{1,1}-g_{2,1}) B_1$, and then decodes $u_{1,2} A_2 +   A_3$.
 Since $ u_{1,2}$ and $g_{1,2}  $ are different elements in $[0,1]^{\rm T}$, it can be seen that user $1$ can recover $A_2$ and $A_3$ from $g_{1,2} A_2  +   A_3$ and $u_{1,2} A_2 +   A_3$. Hence, user $1$ can recover message $A$. We then focus on user $2$ who caches $A_2, B_2$ and desires message $B$. From $ A_{2,2}^{[\bm{\theta}]}$ and its cached content $B_2$, user $2$   decodes $g_{2,1} B_1  +  B_3$.
From $A_1^{[\bm{\theta}]}-A_{2,1}^{[\bm{\theta}]}$, user $2$   decodes
$(u_{1,2}-g_{1,2}) A_2 + v_{1,1} B_1 +g_{2,2} B_2 +   B_3  $, and then decodes $v_{1,1} B_1   +   B_3  $.
Since  $v_{1,1}$ and $g_{2,1}$ are different elements in $[0,1]^{\rm T}$, it can be seen that user $2$ can recover $B_1$ and $B_3$ from $g_{2,1} B_1  +  B_3$ and$v_{1,1} B_1   +   B_3  $. Hence, user $2$ can recover message $B$.

\emph{Privacy:}
Intuitively,  from the viewpoint of DB 1  whose sent linear combination is
$
g_{1,1} A_1+  u_{1,2} A_2+    A_3 + v_{1,1} B_1 + g_{2,2}B_2 +   B_3,
$
the coefficients $g_{1,1}, u_{1,2}, v_{1,1}, g_{2,2} $ are  chosen randomly and independently from the rows of $\mathbf{Y}^{\prime}_2$. Thus the sent linear combination  is independent of the demand vector.
 From the viewpoint of DB 2 whose sent linear combinations are $g_{1,1} A_1 + g_{1,2} A_2 + A_3$ and
$ g_{2,1} B_1 + g_{2,2} B_2 + B_3 $,
the coefficients  $g_{1,1}, g_{1,2} ,  g_{2,1} , g_{2,2}$  are  chosen randomly and independently from the rows of  $\mathbf{Y}^{\prime}_2$. Thus the sent linear combinations are independent of the demand vector. Therefore, each  DB cannot get any information about the demand vector from its sent linear combinations and the user cache.
\emph{Performance:} The achieved load is $R=\frac{D}{L}=1$.
 \hfill $\lozenge$
\end{example}
}

%================================================================================================
\section{Proof of Theorem \ref{theorem 4}: Description of the Product Design}
\label{sec: theorem 4}
In this section, we present a general scheme for arbitrary system parameters, called the {\em Product Design}, which is inspired by
both coded caching \cite{maddah2014fundamental} and the SJ  PIR schemes \cite{sun2017capacity} \footnote{{ In fact, any capacity-achieving PIR scheme can be used to combine with the linear caching code to produce a corresponding product design.}} and enjoys combined coding gain. By comparing with the already established converse bounds for coded caching in \cite{ghasemi2017improved}, we show that the PD is optimal within a factor $8$ in general as indicated in Corollary \ref{corollary: product design opt in high cache regime}.

\if{0}
\begin{example} (Achievability of $(1,\frac{7}{4})$)
Consider the MuPIR problem with $K=3$ messages, $N=2$ DBs and  $K_{\rm u}=3$ users with cache memory $M=1$ (therefore $t=\frac{K_{\rm u}M}{K}=1$). Let $W_1=A,W_2=B$ and $W_3=C$ denote the three messages. Each message is assumed to have $L=24$ bits. The cache placement and private delivery phases are described as follows.

\emph{1) Cache placement:} The Maddah-Ali-Niesen (MAN) cache placment \cite{maddah2014fundamental} is used. More specifically, each message is split into three packets each containing 8 bits, i.e., $A=(A_1,A_2,A_3),B=(B_1,B_2, B_3)$ and $C=(C_1,C_2,C_3)$. The cache placement is
\begin{eqnarray}
Z_1=\{A_1,B_1,C_1\},\quad
Z_2=\{A_2,B_2,C_2\},\quad
Z_3=\{A_3,B_3,C_3\},
\end{eqnarray}
from which it can be seen that the memory constraint is satisfied.

\emph{2) Private delivery:} Suppose the user demands are $\bm{\theta}=(\theta_1,\theta_2,\theta_3)=(1,2,3)$. We first construct three different coded messages $\{X_{\mathcal{S}}^{[\bm{\theta}]}\triangleq(A_{1,\mathcal{S}}^{[\bm{\theta}]}, A_{2,\mathcal{S}}^{[\bm{\theta}]}):\mathcal{S}\subseteq [3],|\mathcal{S}|=2\}$ each being useful to a subset of two users in $\mathcal{S}$. $A_{1,\mathcal{S}}^{[\bm{\theta}]}$ and $A_{2,\mathcal{S}}^{[\bm{\theta}]}$ represents the answers from DB 1 and 2 respectively.

The first coded message is $X_{\{1,2\}}^{[\bm{\theta}]}=(A_{1,\{1,2\}}^{[\bm{\theta}]},A_{2,\{1,2\}}^{[\bm{\theta}]})$ in which
\be\label{eq: X12 1}
A_{1,\{1,2\}}^{[\bm{\theta}]} = A_1^{[\theta_1]}(A_2,B_2,C_2) + A_1^{[\theta_2]}(A_1,B_1,C_1),\ee
\be \label{eq: X12 2}
A_{2,\{1,2\}}^{[\bm{\theta}]} = A_2^{[\theta_1]}(A_2,B_2,C_2) + A_2^{[\theta_2]}(A_1,B_1,C_1),
\ee
where the code components $A_1^{[\theta_1]}(A_2,B_2,C_2)$ and $A_2^{[\theta_1]}(A_2,B_2,C_2)$ represents the answer from DB 1 and DB 2 respectively in the  SJ  PIR scheme when the messages are (First message, second message, third message$)=(A_2,B_2,C_2$) and the user demands $A_2$. The meaning of $A_1^{[\theta_2]}(A_1,B_1,C_1)$ and $A_2^{[\theta_2]}(A_1,B_1,C_1)$ follows similarly. More specifically, let  $A_i=(A_i^1,A_i^2,\cdots,A_i^8)$, $B_i=(B_i^1,B_i^2,\cdots,B_i^8)$ and $C_i=(C_i^1,C_i^2,\cdots,C_i^8),\forall i\in[2]$ be six independent random permutations of the bits of the packets $A_i,B_i,C_i,i\in[2]$. Then the answers from the two DBs are constructed as
\vspace{+0.5cm}
\begin{center}
\begin{tabular}{|c|c|}
\hline
 $A_{1,\{1,2\}}^{[\bm{\theta}]}$ & $A_{2,\{1,2\}}^{[\bm{\theta}]}$ \\\hline
 $A_2^1 + A_1^1$   &   $A_2^2   + A_1^2$\\
$B_2^1 + B_1^1$  & $B_2^2 + B_1^2$\\
 $C_2^1+ C_1^1 $  & $C_2^2+ C_1^2 $\\
$ A_2^3 + B_2^3  + A_1^2+ B_1^2 $  & $ A_2^5 + B_2^1+ A_1^1 + B_1^5  $\\
$ A_2^4 + C_2^2 + A_1^3+ C_1^3 $  & $ A_2^6+ C_2^1+ A_1^4 + C_1^4 $\\
$ B_2^3 + C_2^3 + B_1^4 + C_1^2 $  & $ B_2^4 + C_2^4+ B_1^6+C_1^1$ \\
 $A_2^7 + B_2^4 + C_2^4  + A_1^4  + B_1^7+ C_1^4 $ &  $A_2^8 + B_2^3 + C_2^3  + A_1^3+ B_1^8 + C_1^3 $\\\hline
\end{tabular}
\end{center}
\vspace{+0.5cm}
Note that from the above construction, we can see that in the summations of (\ref{eq: X12 1}) and (\ref{eq: X12 2}), the ordering of the message bits in the query to these answers (and the corresponding DBs) is preserved compared to the  SJ  PIR scheme. Otherwise the DBs may infer some information about the user demands if the orderings of the message bits for different $\bm{\theta}$ are different, which renders the scheme non-private.

The second coded message is
$X_{\{1,3\}}^{[\bm{\theta}]}=(A_{1,\{1,3\}}^{[\bm{\theta}]},A_{2,\{1,3\}}^{[\bm{\theta}]})$ in which
\begin{eqnarray}
A_{1,\{1,3\}}^{[\bm{\theta}]} &=& A_1^{[\theta_1]}(A_3,B_3,C_3) + A_1^{[\theta_3]}(A_1,B_1,C_1), \\
A_{2,\{1,3\}}^{[\bm{\theta}]} &=& A_2^{[\theta_1]}(A_3,B_3,C_3) + A_2^{[\theta_3]}(A_1,B_1,C_1).
\end{eqnarray}
Using a another set of independent random permutations of the bits of the packets $A_i,B_i,C_i,\forall i\in \{1,3\}$, the answers form the two DBs are constructed as
\vspace{+0.5cm}
\begin{center}
\begin{tabular}{|c|c|}
\hline
 $A_{1,\{1,3\}}^{[\bm{\theta}]}$ & $A_{2,\{1,3\}}^{[\bm{\theta}]}$ \\\hline
 $A_3^1 + A_1^1$   &   $A_3^2   + A_1^2$\\
$B_3^1 + B_1^1$  & $B_3^2 + B_1^2$\\
 $C_3^1+ C_1^1 $  & $C_3^2+C_1^2 $\\
$ A_3^3 + B_3^2 + A_1^3 + B_1^3 $  & $ A_3^5 + B_3^1+ A_1^4 + B_1^4 $\\
$ A_3^4 + C_3^2+ A_1^2+ C_1^3 $  & $ A_3^6+ C_3^1+ A_1^1 + C_1^5 $\\
$ B_3^3 + C_3^3 + B_1^2 + C_1^4 $  & $ B_3^4 + C_3^4+ B_1^1+ C_1^6$ \\
 $A_3^7+ B_3^4 +C_3^4+ A_1^4 + B_1^4 + C_1^7 $ &  $A_3^8 + B_3^3 + C_3^3 + A_1^3+ B_1^3 + C_1^8 $\\\hline
\end{tabular}
\end{center}
\vspace{+0.5cm}
The third coded message is $X_{\{2,3\}}^{[\bm{\theta}]}=(A_{1,\{2,3\}}^{[\bm{\theta}]},A_{2,\{2,3\}}^{[\bm{\theta}]})$ in which
\begin{eqnarray}
A_{1,\{2,3\}}^{[\bm{\theta}]} &=& A_1^{[\theta_2]}(A_3,B_3,C_3) + A_1^{[\theta_3]}(A_2,B_2,C_2), \\
A_{2,\{2,3\}}^{[\bm{\theta}]} &=& A_2^{[\theta_2]}(A_3,B_3,C_3) + A_2^{[\theta_3]}(A_2,B_2,C_2).
\end{eqnarray}
Applying a set of independent random permutations of the bits of the message packets $A_i,B_i,C_i,\forall i\in\{2,3\}$, the answers can be constructed as
\vspace{+0.5cm}
\begin{center}
\begin{tabular}{|c|c|}
\hline
 $A_{1,\{2,3\}}^{[\bm{\theta}]}$ & $A_{2,\{2,3\}}^{[\bm{\theta}]}$ \\\hline
 $A_3^1 + A_2^1$   &   $A_3^2   +  A_2^2$\\
$B_3^1 + B_2^1$  & $B_3^2 + B_2^2$\\
 $C_3^1 +  C_2^1 $  & $C_3^2 +  C_2^2 $\\
$ A_3^2  +  B_3^3 + A_2^3  +  B_2^3 $  & $ A_3^1  +  B_3^5 + A_2^4  +  B_2^4 $\\
$ A_3^3 + C_3^3  +  A_2^2 +  C_2^3 $  & $ A_3^4  +  C_3^4 +  A_2^1  +  C_2^5 $\\
$ B_3^4  +  C_3^2 +  B_2^2  +  C_2^4 $  & $ B_3^6 +  C_3^1 +  B_2^1 +  C_2^6$ \\
 $A_3^4 +  B_3^7 +  C_3^4  +  A_2^4  +  B_2^4  +  A_2^7 $ &  $A_3^3 + B_3^8  +  C_3^3  +  A_2^3  +  B_2^3  +  C_2^8 $\\ \hline
\end{tabular}
\end{center}
\vspace{+0.5cm}
Note that for each coded message $X_{\mathcal{S}}^{[\bm{\theta}]}$, a set of independent random permutations (not known to the DBs) are employed to the bits of the involved packets, which is key to privacy.

In the private delivery phase, the users download all the three coded messages from the DBs. We next verify the correctness (i.e., decodability) and privacy of the scheme.

\emph{Correctness:} Let us look at $X_{\{1,2\}}^{[\bm{\theta}]}$ first. Since the packets have been cached by user 1, user 1 can remove the interferences $A_1^{[\theta_2]}(A_1,B_1,C_1),A_2^{[\theta_2]}(A_1,B_1,C_1) $ from (\ref{eq: X12 1}) and (\ref{eq: X12 2}) to obtain the desired coded components $A_1^{[\theta_1]}(A_2,B_2,C_2) $ and $A_2^{[\theta_1]}(A_2,B_2,C_2)$. By the decodability of the  SJ  PIR scheme, user 1 can correctly decode the desired packet $A_2$ from $A_1^{[\theta_1]}(A_2,B_2,C_2) $ and $A_2^{[\theta_1]}(A_2,B_2.C_2)$; Also because the packets $A_2,B_2$ are already cached by user 2, user 2 can remove the interferences $A_1^{[\theta_1]}(A_2,B_2,C_2) $ and $A_2^{[\theta_1]}(A_2,B_2,C_2)$ and obtain the desired code components $A_1^{[\theta_2]}(A_1,B_1,C_1) $ and $A_2^{[\theta_2]}(A_1,B_1,C_1)$, from which the packet $A_1$ can be decoded. Following a similar decoding process, it can be easily seen that from $X_{\{1,3\}}^{[\bm{\theta}]}$, user 1 and 3 can decode $A_3$ and $B_1$ respectively, and from $X_{\{2,3\}}^{[\bm{\theta}]}$, user 2 and 3 can decode $A_3$ and $B_2$ respectively. As a result,the three users can correctly decode their desired messages.

\emph{Privacy:} First note that regardless of the user demands $(\theta_1,\theta_2,\theta_3)\in[2]^3$, three coded messages are downloaded from the DBs. So the DBs can not distinguish different user demands by simply observing the traffic load. Second, each component $A_{n,\mathcal{S}}^{[\bm{\theta}]},n=1,2$ of the coded message $X_{\mathcal{S}}^{[\bm{\theta}]},\forall\mathcal{S}\subseteq [3],|\mathcal{S}|=2$ is independent of the demands of the users in $\mathcal{S}$. The reason is explained as follows. WLOG, we show that both $A_{1,\{1,2\}}^{[\bm{\theta}]}$ and $A_{2,\{1,2\}}^{[\bm{\theta}]}$ are independent of $\theta_1$ and $\theta_2$. By the privacy of the  SJ  scheme, both $A_1^{[\theta_1]}(A_2,B_2)$ and $A_2^{[\theta_1]}(A_2,B_2)$ are independent of $\theta_1$. Also, both $A_1^{[\theta_2]}(A_1,B_1,C_1)$ and $A_2^{[\theta_2]}(A_1,B_1,C_1)$ are independent of $\theta_2$. Therefore, both $A_{1,\{1,2\}}^{[\bm{\theta}]}$ and $A_{2,\{1,2\}}^{[\bm{\theta}]}$ are independent of $\theta_1$ and $\theta_2$. Similarly, both $A_{1,\{1,3\}}^{[\bm{\theta}]}$ and $A_{2,\{1,3\}}^{[\bm{\theta}]}$ are independent of $\theta_1$ and $\theta_3$, and both $A_{1,\{2,3\}}^{[\bm{\theta}]}$ and $A_{2,\{2,3\}}^{[\bm{\theta}]}$ are independent of $\theta_2$ and $\theta_3$. Moreover, since for each coded message $X_{\mathcal{S}}^{[\bm{\theta}]},\forall \mathcal{S}\subseteq [3],|\mathcal{S}|=2$, a set of independent random permutations are applied to the corresponding packet bits, these coded messages are independent of each other. As a result, the answer from each DB $n\in[2]$, i.e., $\{A_{n,\mathcal{S}}^{[\bm{\theta}]}:\mathcal{S}\subseteq [3],|\mathcal{S}|=2\}$ is independent of the user demands $(\theta_1,\theta_2,\theta_3)$. As a result, the scheme is private.

\emph{Performance:} Since $D=42$ bits are downloaded in total, the achieved load is $R=\frac{D}{L}=\frac{7}{4}$, which is better than the naive design with load $K-M=2$.
\hfill $\lozenge$
\end{example}
\fi   % end if{0}

%----------------------------------------------------------------------------
For general system parameters $K,K_{\rm u}$ and $N\ge 2$, we assume that $t=\frac{K_{\rm u}M}{K}\in[1:K_{\rm u}]$. Each message is assumed to have $L=\binom{K_{\rm u}}{t}N^K$ bits. The cache placement and private delivery phases are %formally
described as follows.

\emph{1) Cache placement:} The MAN cache placement is used.  More specifically,  each message $W_k,k\in[K]$ is split into $\binom{K_{\rm u}}{t}$ disjoint and equal-size packets, i.e., $W_k\triangleq\{W_{k,\mathcal{T}}:\mathcal{T}\subseteq [K_{\rm u}],|\mathcal{T}|=t\}$. Therefore, each packet consists of $\frac{L}{\binom{K_{\rm u}}{t}}=N^K$ bits. Each user $u\in[K_{\rm u}]$ caches all the message packets $W_{k,\mathcal{T}}$ such that $u\in\mathcal{T}$, i.e.,
\be
Z_u=\left\{W_{k,\mathcal{T}}:\mathcal{T}\subseteq [K_{\rm u}],|\mathcal{T}|=t,u\in\mathcal{T},\forall k\in[K]   \right\}.
\ee
Therefore, each user stores $KL\frac{\binom{K_{\rm u}-1}{t-1}}{\binom{K_{\rm u}}{t}}=ML$ bits, satisfying the cache memory constraint.

\emph{2) Private Delivery:} Suppose the user demands are $\bm{\theta}=(\theta_1,\theta_2,\cdots,\theta_{K_{\rm u}})\in[K]^{K_{\rm u}}$. We first construct $\binom{K_{\rm u}}{t+1}$ different coded messages
\be
\left\{X_{\mathcal{S}}^{[\bm{\theta}]} \eqdef \left(A_{1,\mathcal{S}}^{[\bm{\theta}]}, A_{2,\mathcal{S}}^{[\bm{\theta}]},\cdots,A_{N,\mathcal{S}}^{[\bm{\theta}]}\right):\mathcal{S}\subseteq [K_{\rm u}],|\mathcal{S}|=t+1\right\},
\ee
each of which being useful to a subset of $t+1$ users in $\mathcal{S}$. The $n$-th component of $X_{\mathcal{S}}^{[\bm{\theta}]}$,  $A_{n,\mathcal{S}}^{[\bm{\theta}]}$, represents the answer from DB $n$. For each %a specific set of users
$\mathcal{S}$, the components of the coded message $X_{\mathcal{S}}^{[\bm{\theta}]}$ are constructed according to the user demands as
\be
\label{Eq: product design summation}
A_{n,\mathcal{S}}^{[\bm{\theta}]}=\sum_{u\in\mathcal{S}}A_{n}^{[\theta_u]}\left( W_{1:K,\mathcal{S}\backslash\{u\}}\right), \forall n\in[N],
\ee
where $W_{1:K,\mathcal{S}\backslash\{u\}}\triangleq\left\{W_{1,\mathcal{S}\backslash \{u\}},W_{2,\mathcal{S}\backslash \{u\}},\cdots,W_{K,\mathcal{S}\backslash \{u\}}\right\}$. The term $A_{n}^{[\theta_u]}( W_{1:K,\mathcal{S}\backslash\{u\}})$ in the summation of (\ref{Eq: product design summation}) represents the answer from DB $n$ in the  SJ  PIR scheme for the single-user PIR problem when the messages are (First message, second message, $\cdots$, $K$-th message) $=(W_{1,\mathcal{S}\backslash \{u\}},W_{2,\mathcal{S}\backslash \{u\}},\cdots,W_{K,\mathcal{S}\backslash \{u\}})$ and the user $u$ demands
$W_{\theta_u,\mathcal{S}\backslash\{u\}}$. To make the scheme private, the users employ a randomly and uniformly distributed permutation (not known to the DBs) of the $N^K$ bits for each packet $W_{\theta_u,\mathcal{S}\backslash\{u\}}$. For different coded messages $X_{\mathcal{S}}^{[\bm{\theta}]}$, the set of the random bit permutations of the corresponding packets are also independent from each other. To ensure demand privacy, the ordering of the message bits in the query to the corresponding answer is preserved as in the  SJ  PIR scheme in the summation of (\ref{Eq: product design summation}). Next  we show that the proposed PD is both correct and private.

\emph{Correctness:} We show that for any user $u\in[K_{\rm u}]$, it can correctly recover its desired message $W_{\theta_{u}}$ from all the answers received. Since the packets {$\{W_{k,\mathcal{T}}:|\mathcal{T}|=t,u\in\mathcal{T},\forall k\in[K]\}$} are cached by user $u$ in the placement phase, it needs to recover the packets $\{W_{\theta_u,\mathcal{T}}:|\mathcal{T}|=t,u\notin\mathcal{T}\}$.
For each $n\in{[N]},\mathcal{S}\subseteq [K_{\rm u}]$ such that $|\mathcal{S}|=t+1, u\in\mathcal{S}$, we can write (\ref{Eq: product design summation}) as
\be
\label{Eq: product design summation-extended}
A_{n,\mathcal{S}}^{[\bm{\theta}]}=A_{n}^{[\theta_u]}\left( W_{1:K,\mathcal{S}\backslash\{u\}}\right)+\sum_{u'\in\mathcal{S}\backslash \{u\}}A_{n}^{[\theta_{u'}]}\left( W_{1:K,\mathcal{S}\backslash\{u'\}}\right),
\ee
from which user $u$ can decode the desired term $A_{n}^{[\theta_u]}\left( W_{1:K,\mathcal{S}\backslash\{u\}}\right)$ since all the packets $\{W_{1:K,\mathcal{S}\backslash\{u'\}}:u'\ne u\}$ are cached by user $u$ because $u\in \mathcal{S}\backslash \{u'\}$. Therefore, user $u$ obtains a set of desired answers $\{A_{n}^{[\theta_u]}\left( W_{1:K,\mathcal{S}\backslash\{u\}}\right):\forall n\in[N]\}$ from which the desire packet $W_{\theta_{u},\mathcal{S}\backslash \{u\}}$ can be decoded  due to the decodability of the  SJ  PIR scheme. Going through all different such user subsets $\mathcal{S}$, user $u$ can decode all the $\binom{K_{\rm u}-1}{t}$ desired packets. As a result, user $u$ can correctly recover its desired message $W_{\theta_u}$.

\emph{Privacy:}
It can be seen that each $A_{n,\mathcal{S}}^{[\bm{\theta}]},\forall n\in[N]$ of the coded message $X_{\mathcal{S}}^{[\bm{\theta}]}$ is independent of the demands of the users in $\mathcal{S}$ and that of the users in $[K_{\rm u}]\backslash\mathcal{S}$ from the perspective of each individual DB. The reason is explained as follows. For any user $u\in\mathcal{S}$, we see that the first term in  (\ref{Eq: product design summation-extended}), i.e.,  $A_{n}^{[\theta_u]}\left( W_{1:K,\mathcal{S}\backslash\{u\}}\right),\forall n\in[N]$ is independent of $\theta_u$ by the privacy of the  SJ  PIR scheme. Also, each item in the summation of the second term in (\ref{Eq: product design summation-extended}) is independent of $\theta_u$ because for each $A_{n}^{[\theta_{u'}]}\left( W_{1:K,\mathcal{S}\backslash\{u'\}}\right)$, a set of random and independent permutations are employed to the bits of the set of packets  $\{W_{k,\mathcal{S}\backslash\{u'\}}:\forall k\in[K]\}$. Therefore, $A_{n,\mathcal{S}}^{[\bm{\theta}]}$ is independent of the demands of the users in $\mathcal{S}$. Moreover, for $\mathcal{S}$ where $u\notin \mathcal{S}$, due to the employment of the random and independent permutations on the packet bits, $A_{n,\mathcal{S}}^{[\bm{\theta}]}$ is independent of the demands of the users in $[K_{\rm u}]\backslash\mathcal{S}$. As a result, $A_{n,\mathcal{S}}^{[\bm{\theta}]}$ is independent of $\bm{\theta}$ from DB $n$'s perspective for any $\mathcal{S}\subseteq [K_{\rm u}],|\mathcal{S}|=t+1$, which completes the privacy proof for the PD.

\emph{Performance:} By the  SJ  PIR scheme, each
$X_{\mathcal{S}}^{[\bm{\theta}]}$ has $\left( 1+\frac{1}{N}+\cdots+\frac{1}{N^{K-1}}\right)N^K$ bits. Therefore, $D=\binom{K_{\rm u}}{t+1}\left( 1+\frac{1}{N}+ \cdots + \frac{1}{N^K}\right)N^K$ bits are downloaded from the DBs in total. As a result, the achieved load is
$\widehat{R}(M) =\frac{D}{L}=\frac{K_{\rm u}-t}{t+1}\left( 1+\frac{1}{N}+\cdots+\frac{1}{N^{K-1}}\right)$.

%==================================================================================================
\section{Discussion: MuPIR with Distinct Demands}
\label{sec: Discussions}
In this section, we consider an interesting scenario where the users have distinct demands. Recall that $R_{\rm d}^{\star}$ denotes the minimum load. We obtain the following theorem.
\begin{theorem}
\label{theorem 3}
For the cache-aided MuPIR problem with $K=2$ messages, $K_{\rm u}=2$ users and $N=2$ DBs, where the users demand distinct messages in a uniform manner, the optimal memory-load trade-off is characterized as
\begin{equation}\label{eqn: optimal distinct demands}
  R_{\rm d}^{\star}(M) =
    \begin{cases}
      2(1-M) ,& 0\le M\le \frac{1}{3}\\
      \frac{5}{3}-M ,  & \frac{1}{3}\le M\le \frac{2}{3}\\
      \frac{3(2-M)}{4}, &\frac{2}{3}\le M\le 2
    \end{cases}.
\end{equation}
\end{theorem}
\begin{IEEEproof}
For achievability, we show that the memory-load pairs $(\frac{1}{3},\frac{4}{3})$ and $(\frac{2}{3},1)$ are achievable using the idea of CIA in Section~\ref{sec: achievability}. Together with the two trivial pairs $(0,2)$ and $(2,0)$, we obtain four corner points. By memory sharing among these corner points, the load of Theorem \ref{theorem 3} can be achieved. For the converse, when $M\le\frac{1}{3}$, the load of $R_{\rm d}(M)=2(1-M)$ coincides with the caching bound without demand privacy and hence is optimal. When $M\ge \frac{2}{3}$, the load $R_{\rm d}(M)=\frac{3(2-M)}{4}$ is optimal since it coincides with the single-user cache-aided PIR bound of \cite{tandon2017capacity}. A novel converse bound is derived for the cache memory regime $\frac{1}{3}\le M\le \frac{2}{3}$ to show the optimality of $R(M)=\frac{5}{3}-M$ when the users have distinct demands.
\end{IEEEproof}

\begin{remark}
It can be see that the load in (\ref{eqn: optimal distinct demands})  is lower than  the one in (\ref{theorem 1}) achieved by the CIA based scheme. Thus  the optimal load under the constraint of distinct demands can be strictly lower  than the  optimal load without such constraint (See Fig.~\ref{fig: load_2DBs}).
\end{remark}

\subsection{Achievability}
\label{sec: achievability}
First, we consider the achievability of the memory-load pair $(\frac{1}{3},\frac{4}{3})$.
Assume that the users have distinct demands, i.e., the demand vector $\bm{\theta}=(\theta_1,\theta_2)$ can only be $(1,2)$ or $(2,1)$.
Let $A_{1,1}$ and $A_{1,2}$ be two different answers from DB 1, and let $A_{2,1}$ and $A_{2,2}$ be two different answers from DB 2.
Assume that each message contains $L=3$ bits,
i.e., $W_1=(A_1,A_2,A_3),W_2=(B_1,B_2,B_3)$. The cache placement is
$Z_1=\{A_1+B_1\}, Z_2=\{A_2+ B_2\}$
and the answers are %constructed as
$A_{1,1}=(A_3,B_1+B_2 + B_3),
A_{1,2}=(A_1+ A_2+A_3,B_3),
A_{2,1}=(A_2+ A_3,B_2+ B_3),
A_{2,2}=(A_1+ A_3,B_1+ B_3)$.

The private delivery scheme is that the users randomly choose $A_{1,1}$ or $A_{1,2}$ to request from DB 1 with equal probabilities. We then consider the following two cases.
When $(\theta_1,\theta_2)=(1,2)$, if $A_{1,1}$ is chosen, then go to DB 2 to download $A_{2,1}$. Otherwise, if $A_{1,2}$ is chosen, go to DB 2 to download $A_{2,2}$.
When  $(\theta_1,\theta_2)=(2,1)$, if $A_{1,1}$ is chosen, then go to DB 2 to download $A_{2,2}$. Otherwise if $A_{1,2}$ is chosen, go to DB 2 to download $A_{2,1}$.

For the correctness of this scheme, one can check that
$(A_{1,1},A_{2,1},Z_1)\to W_1$, \footnote{{ The notation $(A_{1,1},A_{2,1},Z_1)\to W_1$ means that the message $ W_1$ can be recovered from $A_{1,1},A_{2,1}$, and $Z_1$. Other notations follow similarly.}} $(A_{1,1},A_{2,1},Z_2)\to W_2$
 $(A_{1,2},A_{2,2},Z_1)\to W_1$, $(A_{1,2},A_{2,2},Z_2)\to W_2$,
 $(A_{1,1},A_{2,2},Z_1)\to W_2$, $(A_{1,1},A_{2,2},Z_2)\to W_1$,
$(A_{1,2},A_{2,1},Z_1)\to W_2$, and $(A_{1,2},A_{2,1},Z_2)\to W_1$.
Therefore, all users can decode their desired messages. For privacy of this scheme, note that the answer from DB 1 is equally likely to be $A_{1,1}$ or $A_{1,2}$, and the answer from DB 2 is also equally likely to be $A_{2,1}$ or $A_{2,2}$. Therefore, we have
$P\left(\bm{\theta}=(1,2)\right)=P((A_{1,1},A_{2,1}))+{P}((A_{1,2},A_{2,2}))=\frac{1}{2}$ and
$P(\bm{\theta}=(2,1))=P((A_{1,1},A_{2,2}))+P((A_{1,2},A_{2,1}))=\frac{1}{2}$ such that the privacy constraint (\ref{eqn: privacy}) is satisfied (for distinct demands). Since $D=4$ bits are downloaded in total, the achieved load is $R_{\rm d}=\frac{D}{L}=\frac{4}{3}$.

Second, we consider the achievability of the memory-load pair  $(\frac{2}{3},1)$. Assume that each message has $L=3$ bits, i.e., $W_1=(A_1,A_2,A_3),W_2=(B_1,B_2,B_3)$. The cache placement is
$Z_1=\{A_1,B_1\}, Z_2=\{A_2,B_2\}$
and the answers from the DBs are %constructed as
$A_{1,1}=(A_3+ B_3 + B_1+ B_2)$,
 $A_{1,2}=(A_3+ B_3+ A_1+A_2)$, $
A_{2,1}=(A_2+ A_3,B_2+ B_3)$ and $
A_{2,2}=(A_1+ A_3,B_1+B_3)$.

The private delivery phase works similarly to the above corner point $(\frac{1}{3},\frac{4}{3})$. The correctness of this scheme can be checked straightforwardly using the above transmitted codewords and cached information.
The privacy argument is similar to the previous case, i.e., from each DB's perspective, the demand vector $\bm{\theta}$ is equally likely to be $(1,2)$ or $(2,1)$. Since $D=3$ bits are downloaded in total, the achieved load is $R_{\rm d}=\frac{D}{L}=1$.

\vspace{-0.5cm}
\subsection{Converse}
The converse curve consists of three piece-wise linear segments corresponding to different cache memory regimes $0\le M\le \frac{1}{3}$, $\frac{1}{3}\le M\le \frac{2}{3}$, and $\frac{2}{3}\le M\le 2$. We prove the converse for each  segment respectively.

\subsubsection{$0\le M\le \frac{1}{3}$} In this regime, the cut-set bound {without   privacy constriant} is tight (see Corollary~\ref{corollary: theorem 1 optimality}).

\subsubsection{$\frac{1}{3}\le M\le \frac{2}{3}$}
Let $A_{1,1}=A_{1,1}^{[(1,2)]}=A_{1,1}^{[(2,1)]}$ be an answer of DB 1 and let $A_{2,1}=A_{2,1}^{[(1,2)]}=A_{2,1}^{[(2,1)]}$ be an answer of DB 2. It is clear that the message $W_1$ can be recovered from $\{A_{1,1}^{[(1,2)]},A_{2,1}^{[(1,2)]},Z_1$\} while $W_2$ can be recovered from $\{A_{1,1}^{[(1,2)]},A_{2,1}^{[(1,2)]},Z_2\}$ for which we use a shorthand notation as $(A_{1,1}^{[(1,2)]},A_{2,1}^{[(1,2)]},Z_1)\to W_1$, $(A_{1,1}^{[(1,2)]},A_{2,1}^{[(1,2)]},Z_2)\to W_2$. For privacy of DB 1, there must exist another answer $A_{2,2}^{[(2,1)]}$ of DB 2 such that the demands $\bm{\theta}=(2,1)$ can be decoded, i.e., $(A_{1,1}^{[(2,1)]},A_{2,2}^{[(2,1)]},Z_1)\to W_2$ and $(A_{1,1}^{[(2,1)]},A_{2,2}^{[(2,1)]},Z_2)\to W_1$. Also, for privacy of DB 2, there must exist another answer $A_{1,2}^{[(2,1)]}$ of DB 1 such that $(A_{1,2}^{[(2,1)]}, A_{2,1}^{[(2,1)]},Z_1)\to W_2$ and $(A_{1,2}^{[(2,1)]}, A_{2,1}^{[(2,1)]},Z_2)\to W_1$. Note that $R_{\rm d}=\left(H(A_{1,i})+H(A_{2,j})\right)/L$ for any index pair $(i,j)\in \{(1,1),(1,2),(2,1)\}$ (Load does not depend on $\bm{\theta}$). Denote $X_{i,j,k}^{[\bm{\theta}]}\triangleq(A_{1,i}^{[\bm{\theta}]},A_{2,j}^{[\bm{\theta}]},Z_k),\forall i(i,j)\in \{(1,1),(1,2),(2,1)\}$. We then have
\begin{align}
& R_{\rm d}(M)L+ 3ML \nonumber\\
& \ge H(X_{1,1,1}^{[(1,2)]})+H(X_{2,1,2}^{[(2,1)]})+H(X_{1,2,2}^{[(2,1)]}) \nonumber\\
&\overset{\textrm{(a)}}{=}  3L+ H(X_{1,1,1}^{[(1,2)]}|W_1)+H(X_{2,1,2}^{[(2,1)]}|W_1)+H(X_{1,2,2}^{[(2,1)]}|W_1) \nonumber\\
& \overset{\textrm{(b)}}{\ge} 3L+ H(X_{1,1,1}^{[(1,2)]}|W_1) + H(Z_2|W_1) + H(A_{2,1}^{[(2,1)]}|W_1,Z_2)
+H(X_{1,2,2}^{[(2,1)]}|W_1)\nonumber \\
& \ge 3L+ H(X_{1,1,1}^{[(1,2)]},Z_2|W_1)  + H(A_{2,1}^{[(2,1)]}|W_1,Z_2)
+H(X_{1,2,2}^{[(2,1)]}|W_1)\nonumber\\
&\overset{\textrm{(c)}}{=}   4L + H(A_{2,1}^{[(2,1)]}|W_1,Z_2)+H(X_{1,2,2}^{[(2,1)]}|W_1)\nonumber\\
&\ge 4L +H(A_{2,1}^{[(2,1)]}|W_1,Z_2)+H(Z_2|W_1) +  H(A_{1,1}^{[(2,1)]}|W_1,Z_2)\nonumber\\
&\ge 4L +H(A_{1,1}^{[(2,1)]},A_{2,1}^{[(2,1)]}|W_1,Z_2)+H(Z_2|W_1) \nonumber\\
&= 4L+ H(A_{1,1}^{[(2,1)]},A_{2,1}^{[(2,1)]},Z_2|W_1)\nonumber \\
&= 4L+ H(A_{1,1}^{[(1,2)]},A_{2,1}^{[(1,2)]},Z_2|W_1)\nonumber\\
&= 5L,
\label{Eq: distinct demand converse}
\end{align}
where (a) is due to $X_{1,1,1}^{[(1,2)]}\to W_1$, $X_{2,1,2}^{[(2,1)]}\to W_1$ and $X_{1,2,2}^{[(2,1)]}\to W_1$; in (b) we used the chain rule and non-negativity of mutual information; (c) is due to the fact that both $W_1$ and $W_2$ can be decoded from $X_{1,1,1}^{[(1,2)]}$ and $Z_2$. (\ref{Eq: distinct demand converse}) implies $R_{\rm d}(M)\ge \frac{5}{3}-M$, which completes the converse proof.

\subsubsection{$\frac{2}{3}\le M \le 2$} In this regime, the achievable load $R_{\rm d}(M)=\frac{3(2-M)}{4}$ coincides with the  single-user cache-aided PIR bound given in \cite{tandon2017capacity} as stated in Corollary \ref{corollary: theorem 1 optimality}. Since increasing the number of users while keeping the user demands private from the DBs can only possibly increase the load, we conclude that $R(M)\ge \frac{3(2-M)}{4}$. This completes the converse proof of Theorem \ref{theorem 3}.

\section{Conclusions and Future Directions}
\label{sec: conclusions}
In this paper, we introduced the problem of cache-aided Multiuser Private Information Retrieval (MuPIR) problem, which generalized the single-user PIR to the case of multiple users.
We provided achievability for the MuPIR problem with two messages, two users and arbitrary number of databases utilizing the novel idea of cache-aided interference alignment (CIA). The proposed scheme is shown to be optimal when the cache placement is uncoded.
For general systems parameters,  inspired by both single-user PIR and coded caching, we proposed a product design
which is order optimal within a factor of 8.
Moreover, when the demands are constrained to be distinct, the optimal memory-load tradeoff is characterized for a system with two messages, two users and two databases.
 Due to the strong connection to both PIR and coded caching, our result on the cache-aided MuPIR problem provides useful insights to understanding the role of side information (i.e., cache) in multiuser and multi-message PIR. Besides the proposed achievability and converse results, the cache-aided MuPIR problem still remains open for arbitrary system parameters in terms of the optimal memory-load trade-off.
For example, utilizing the idea of CIA, we expect more achievability results to come. Also, based on the well-established converse results of coded caching and PIR, a systematic approach to characterize the converse is needed.

\bibliographystyle{IEEEtran}
\bibliography{references_d2d}

\end{document}